\documentclass[12pt]{article}

\usepackage{newtxtext,newtxmath}
\usepackage{graphicx}

\usepackage[letterpaper,margin=1in]{geometry}

\renewenvironment{abstract}
{\quotation}
{\endquotation}

\date{}

\makeatletter
\renewcommand{\fnum@figure}{\textbf{Figure \thefigure}}
\renewcommand{\fnum@table}{\textbf{Table \thetable}}
\makeatother

\usepackage{scicite}

\usepackage{url}

\usepackage{physics}
\usepackage{array}
\usepackage{amsmath}
\usepackage{subfigure}

\def\scititle{
	Metasurface quantum graphs for polarization-independent certified randomness 
}
\title{\bfseries \boldmath \scititle}

\author{
	Xing Lin$^{1,3\dagger}$,
	Wenwen Liu$^{1,3\dagger}$,
	Wange Song$^{1,3,6\dagger}$,
	Sixin Chen$^{1,3}$,
	Shuang Zhang$^{1,2,3,4,5\ast}$\and
	\small$^{1}$New Cornerstone Science Laboratory, Department of Physics, The University of Hong Kong,\\\small Hong Kong, 999077, China.\and
	\small$^{2}$Department of Electrical \& Electronic Engineering, The University of Hong Kong, Hong Kong 999077, China.\and
	\small$^{3}$State Key Laboratory of Optical Quantum Materials, University of Hong Kong, Hong Kong 999077, China.\and
	\small$^{4}$Materials Innovation Institute for Life Sciences and Energy (MILES), HKU-SIRI, Shenzhen 518000, China.\and
	\small$^{5}$Quantum Science Center of Guangdong-Hong Kong-Macao Great Bay Area, 3 Binlang Road, Shenzhen, China\and
	\small$^{6}$National Laboratory of Solid State Microstructures,\and \small Key Laboratory of Intelligent Optical Sensing and Manipulations,\and\small Jiangsu Key Laboratory of Artificial Functional Materials,\and\small Collaborative innovation center of advanced microstructures,\and\small College of Engineering and Applied Sciences, Nanjing University, Nanjing, 210093, China.\and
	\small$^\ast$Corresponding author. Email: shuzhang@hku.hk\and\and
	\small$^\dagger$These authors contributed equally to this work.\and
}

\begin{document} 
	
	% Insert the title and author list
	\maketitle
	
	% Abstract, in bold
	% There are strict length limits, and not all formats have abstracts.
	% Consult the journal instructions to authors for details.
	% Do not cite any references in the abstract.
	\begin{abstract}\bfseries \boldmath
		Quantum random number generators are crucial for information security, yet existing implementations are hindered by bulky optics, device imperfections, and state alignment constraints. Here, using a compact metasurface platform, we achieve polarization-independent certified randomness by establishing a quantitative quantum graph framework. By mapping multi-channel nanophotonic responses onto graph vertices, we assign metasurface measurements an operational information-theoretic meaning, directly connecting weighted graph degrees to conditional min-entropy estimation. The resulting graph-certified measurements yield robust private randomness even under entanglement-based attacks without complex optics. Experimentally, we fabricate 4- and 6-channel dielectric metasurfaces to implement high-dimensional positive operator-valued measure graphs, successfully certifying randomness with heralded single photons and Poissonian light. By integrating complex quantum measurements into a single passive layer, our metasurface quantum graphs provide a scalable route for integrated quantum technologies.
	\end{abstract}

	\section*{Introduction}

	The performance and reliability of our digital, networked society critically depend on access to private, high-quality randomness. In the cryptocurrency sector alone, vulnerabilities arising from flawed pseudo-random entropy sources have led to cumulative losses of hundreds of billions of dollars in recent years \cite{wang2020ecdsa,heninger2012mining}. To meet escalating information security demands, random number generation has shifted from conventional pseudo-random algorithms—fundamentally vulnerable to computational and side-channel attacks—to quantum random number generators (QRNGs) that harness the intrinsic unpredictability of quantum measurements \cite{born1926quantenmechanik}. Such QRNGs are now central to modern cryptography, large-scale scientific simulations, and emerging quantum technologies \cite{herrero2017quantum,acin2016certified}.
	
	Early QRNG implementations, for example those based on measuring photon polarization in a diagonal basis using polarization beam splitters, successfully demonstrated the principle of quantum randomness \cite{jennewein2000fast}. However, unavoidable imperfections in practical devices, together with growing concerns about potential adversarial manipulation of the hardware, have driven the field toward certified quantum randomness using device-independent-type protocols. These include fully device-independent (DI) schemes grounded in Bell inequality violations \cite{pironio2010random,christensen2013detection,bierhorst2018experimentally,liu2018device,liu2021device,liu2025certified}, and semi-device-independent (Semi-DI) approaches based on entropy-uncertainty relations \cite{cao2016source,marangon2017source,avesani2018source,drahi2020certified,avesani2022unbounded,lin2020security,lin2022imperfection,lin2022certified,wang2023provably,lin2023quantum}. Despite their strong security guarantees, both DI and semi-DI frameworks encounter severe scalability challenges: they impose stringent protocol and security requirements and typically rely on bulky, complex linear-optical setups that are difficult to scale to higher-dimensional, multi-channel quantum systems.
	
	Metasurfaces, consisting of ultrathin, lightweight two-dimensional arrays of subwavelength dielectric or metallic meta-atoms, have revolutionized wavefront control by providing precise manipulation of light’s amplitude, phase, polarization, and propagation direction in a compact form \cite{yu2011light,dorrah2022tunable,kuznetsov2016optically}. Unlike traditional bulk optics, such as lenses with wavelength-scale thicknesses, metasurfaces engineer light–matter interactions through carefully designed nanoantenna geometries, resonances, and spatial arrangements, including Pancharatnam–Berry phase gradients that impart polarization-dependent phase shifts. Advances in nanofabrication and subwavelength photonics have driven metasurfaces from classical applications in imaging \cite{khorasaninejad2016metalenses}, holography \cite{rubin2019matrix}, and multifunctional beam steering \cite{arbabi2015dielectric} toward quantum optics, where they have been used to realize photon entanglement \cite{li2020metalens,santiago2022resonant,stav2018quantum}, multi-photon interference \cite{yousef2025metasurface,li2021non}, and quantum state reconstruction \cite{wang2018quantum,solntsev2021metasurfaces}.
	
	Yet, while metasurfaces offer scalability advantages, harnessing them for high-dimensional quantum information tasks, such as device-independent—type QRNGs, introduces new complexity: as the number of photons and output ports grows, it becomes increasingly difficult to model correlations, quantify entropy, and guarantee security against sophisticated attacks, especially under imprecise characterizations on the devices. Graph theory provides an intuitive and powerful toolset for addressing these challenges, building on the success of graph states in fault-tolerant quantum computing \cite{gu2019quantum,bartolucci2023fusion}. Recent work introduced a graph-based representation of metasurfaces as a descriptive and design language for visualizing multichannel Hong-Ou-Mandel interference \cite{yousef2025metasurface}. However, beyond the visualization paradigm, a unified framework that treat metasurface graphs as quantitative information-theoretic objects, thereby bridging high-dimensional quantum information tasks, metasurface photonics, and graph theory, remains lacking.

	Here, we employ quantum graph theory to map the metasurface output ports onto corresponding positive operator-valued measure (POVM) elements and thereby quantify secure, scalable, input-state polarization-independent quantum certified randomness. Our weighted graph framework intuitively reveals the correlations among these POVM elements, connecting weight degrees to conditional min-entropy. This, in turn, enables robust randomness extraction from arbitrary polarization inputs, with multiphoton or mixed Fock-state, across high-dimensional multi-channel outputs, while ensuring security even against eavesdroppers entanglement attacks. By endowing metasurface graphs with an explicit operational interpretation in entropy estimation and security certification, we transform complex high-dimensional measurement architectures--essential for scalable QRNGs--into a unified single-layer metasurface graph representation for quantum correlation analysis and uncertainty quantification. This establishes a direct quantum graph representation that seamlessly integrates metasurface design with entropy quantification--the core metric in quantum information science.

	\section*{Result}
	
	\subsection*{\label{sec:level1}Metasurface POVM graph representation}
	
	Graph-based representations have proven useful for visualizing interference phenomena in metasurfaces \cite{yousef2025metasurface}. We consider a metasurface composed of spatially varying nanostructures that supports $N$ output channels, each associated with a distinct diffraction order whose polarization response is independently engineered by the corresponding super-atom subwavelength grating, as shown in Fig. \ref{fig:1}. The polarization transformation in the $i$th channel is described by a Jones matrix $J_i$. These matrix-valued diffraction orders define a graph $G=(V,E)$, in which vertex $i\in V$ represents the optical path encoded by $J_i$, while edges are defined through the Hilbert-Schmidt inner products $\langle J_i,J_j\rangle$ between vertices $i$ and $j$ \cite{yousef2025metasurface}. 
	
	We extend this graph language into a quantitative framework compatible with quantum information theory by associating the metasurface output ports with POVM elements. The diffraction efficiency depends on the metasurface's nanoanenna geometry, the metagrating arrangement and material properties, enabling tailored polarization manipulation for quantum measurements. To incorporate port-dependent efficiencies, we define the POVM $\{F_i\}_{N+1}$ as $\left\{ t \eta_1 J_1, t \eta_2 J_2, \dots, t \eta_N J_N, I - \sum_{i=1}^N t \eta_i J_i \right\}$, where $\eta_i $ denotes the transmission efficiency of the $i$th channel, and $t>0$ is chosen such that $\sum_{i=1}^{N} t \eta_i J_i \le I$. The last term, representing possible loss, is excluded from analysis when focusing solely on effective responses. Representative POVM graphs are shown in Fig. \ref{fig:2}. In this weighted-graph description, vertex weights encode the relative efficiencies $\eta_i$, thereby providing a more realistic account of experimental imperfections. Fig. \ref{fig:2}(a, b) present several representative POVM graphs together with the corresponding Bloch-sphere projections of their measurement elements, highlighting the underlying geometry in the polarization Hilbert space. For two orthogonal measurement elements, the Hilbert-Schmidt inner product vanishes, $\langle J_i,J_j\rangle=0$, and no edge is formed; accordingly, a two-element POVM yields a disconnected graph. By contrast, when the number of POVM elements exceeds the polarization-space dimension (2 for polarization space), nonorthogonal overlaps become unavoidable, and edges necessarily appear between vertices. These overlaps encode the intrinsic correlations among measurement outcomes and are essential for generating quantum correlations.

	For optimally symmetric POVM frames—such as those achieving informationally complete measurements—the graph exhibits a balanced structure: orthogonal vertices remain isolated (no edges), while uniformly-spaced non-orthogonal pairs are connected by edges of equal weight, preserving central symmetry and maximizing measurement efficiency, as illustrated in Fig. \ref{fig:2}(a). In contrast, deviations from this optimal condition break the symmetry, leading to non-uniform edge weights or even missing edges, as illustrated in Fig. \ref{fig:2}(b). Such asymmetries alter the graph's topology, directly impacting the degree of each vertex and, consequently, the system's performance in quantum measurement applications. To quantify this, we define the weighted degree of a vertex corresponding to $J_{i} $ as
	\begin{eqnarray}
		D_{i}= \frac{1}{\eta_{i}}\sum_{j=1}^{N} \eta_{j} \langle J_{i},J_{j} \rangle  \label{de}
	\end{eqnarray}
	where the summation incorporates the relative efficiencies $\eta_j$ to account for channel imbalances. When the POVM cardinality surpasses that of the computational basis (i.e., overcomplete measurements), non-zero weighted degrees emerge persistently, signaling the presence of non-classical correlations.

	\subsection*{\label{sec:level2}Polarization-independent randomness estimation}
	
	The correlations between POVM elements, as represented in the graph, directly underpin the certification of quantum randomness. For orthogonal POVM elements, such as the horizontal-vertical (H-V) polarization basis, a diagonal polarization input yields maximum randomness due to the balanced outcome probabilities. However, if the input state is aligned with the measurement basis, such as horizontal polarization, no quantum randomness is generated, rendering the randomness polarization-dependent. Metasurfaces overcome this limitation by enabling scalable expansion to additional ports, such as a diagonal (D) polarization port, ensuring certified randomness regardless of the input state (as shown in Fig. \ref{fig:2}(b)). In the minimal randomness scenario, where an H-polarized input aligns with a measurement basis, a metasurface configured with H, V, and D polarization ports yields $\log_2 \frac{3}{2}$ bits of randomness per measurement outcome, demonstrating robust polarization-independent randomness generation.

\begin{figure}[htbp]
		\centering
		\includegraphics[width=5in]{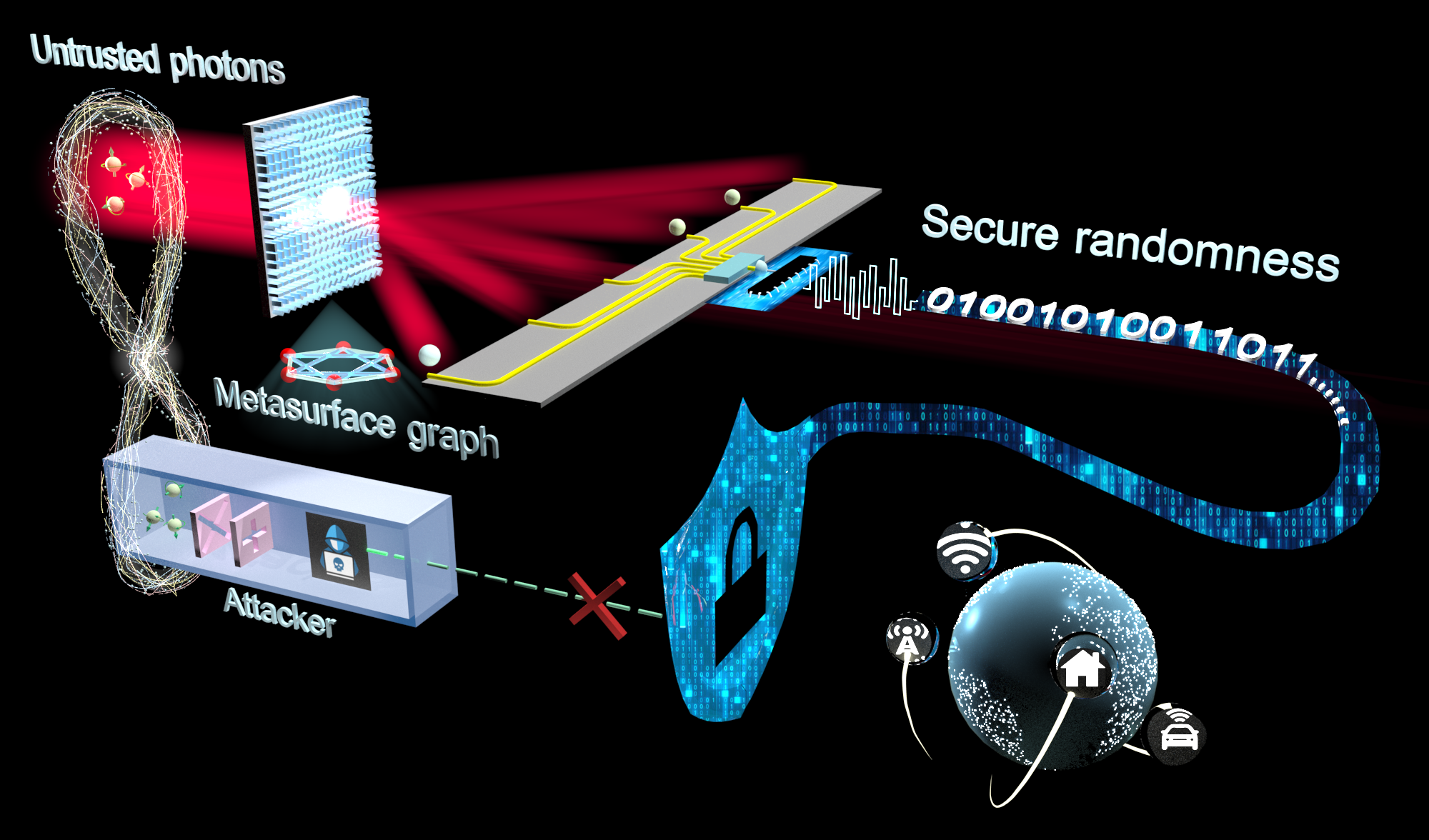}
		\caption{Schematic illustration of a metasurface quantum graph for secure quantum randomness generation. Untrusted input photons may be entangled with an adversary, while the metasurface implements a multichannel quantum measurement encoded by a quantum graph to generate and certify private randomness, thereby preventing adversarial access to the protected information.}
		\label{fig:1}
	\end{figure}

\begin{figure}[htbp]
		\centering
		\includegraphics[width=6.5in]{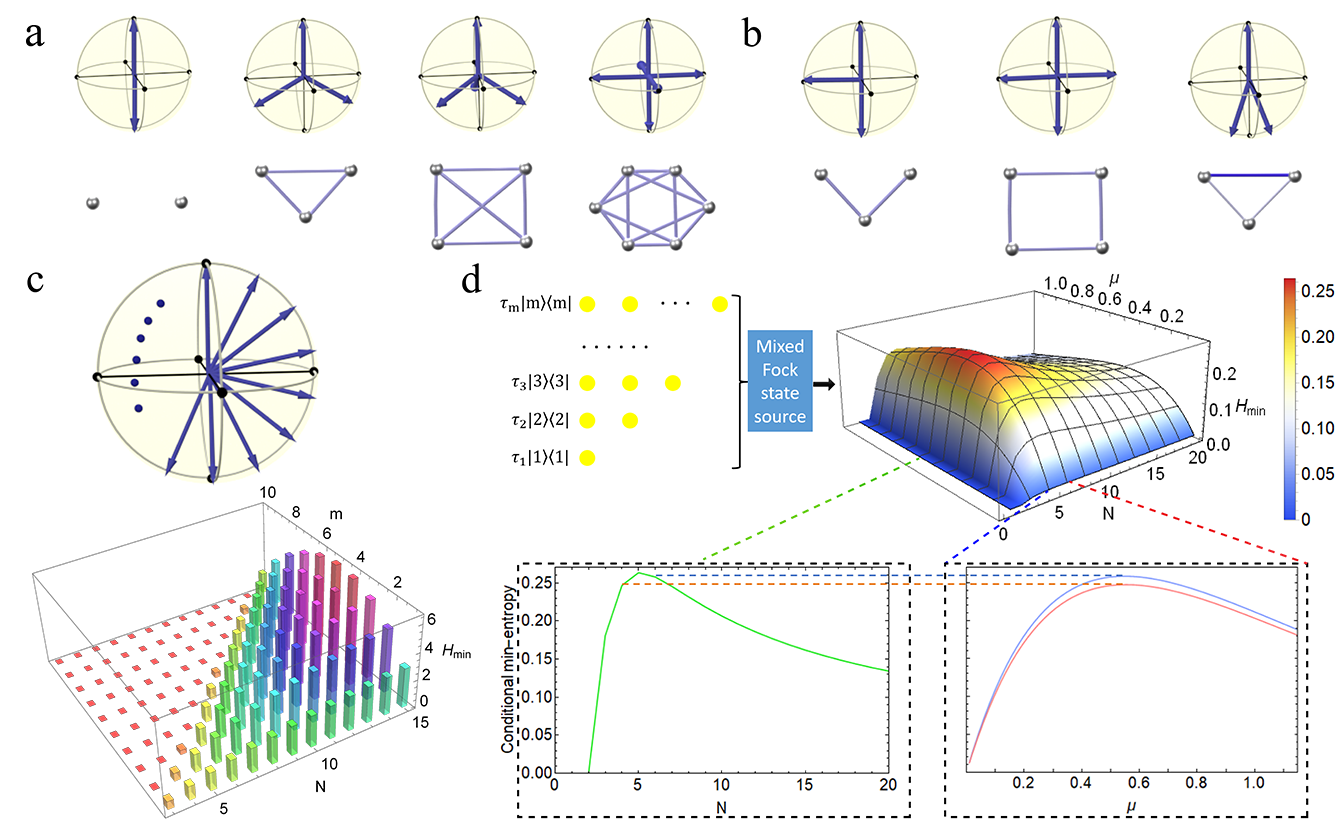}
		\caption{Quantum-graph representations of multi-channel POVMs and conditional min-entropy theory. (a) Optimally symmetric POVM frames (ideal informationally complete configurations) and their associated quantum graphs for 2, 3, 4, and 6 channels, showing uniform edge weights and preserved central symmetry. (b) Examples of non-optimally symmetric POVM frames for 3 and 4 channels, together with their corresponding quantum graphs, exhibiting non-uniform edge weights or missing edges due to broken symmetry and channel imperfections. (c) Conditional min-entropy for a metasurface with $  N  $ output ports uniformly spaced in linear polarization angle across the Bloch plane: top—schematic representation of the measurement bases on the Bloch sphere; bottom—3D point map showing $  H_{\min}(A|E)  $ as a function of channel number $  N  $ and input photon number $  m  $. (d) Conditional min-entropy for realistic light sources with mixed Fock-state statistics: top—3D heatmap for overall secure min-entropy $  H_{\min}(A_{\tau}|E)  $ incorporating worst-case eavesdropping attacks as a function of channel number $  N  $ and the mean photon number $  \mu  $ of the input state; bottom left—dependence on $  N  $ with $  \mu=0.6  $; bottom right—dependence on $  \mu  $ with $  N=  $ 4 or 6. }
		\label{fig:2}
	\end{figure}

	To quantify the certified quantum randomness for an unknown source emitting $m$-photon states $\rho_A^m$ onto an $N$-channel metasurface described by the POVM $\{F_i\}_N$, we consider an entanglement-based attack where an eavesdropper, Eve, holds the purification $\ket{\psi_{AE}}$ of the input state $\rho_A^{m} = \Tr_E[\ketbra{\psi_{AE}}]$ and performs an optimal POVM $\{M_i^E\}_N$ on her subsystem to guess the metasurface outcome $i$, as illustrated in Fig.~\ref{fig:1}. The guessing probability for the final random sequence $A$ is then (see Materials and Methods)
	\begin{eqnarray}
		p_{\text{guess}}(A|\rho_{A}^{m}, \{F_{i}\}_{N})
		&=&\max_{\{\{M^{E}_{i}\}_{N},\ket{\psi_{AE}}\}}\sum_{i}
		\Tr\!\big[F_{i}^{\otimes m}\otimes M^{E}_{i}\ketbra{\psi_{AE}}\big]
		\nonumber\\
		&\leq& \max_{i}\{D_{i}^{-m}\},
	\end{eqnarray}
	The upper bound is proved to be determined by the POVM graph and, in particular, by the minimum weighted degree $D_{i}$ of its vertices. For two orthogonal POVM elements, Eve can exploit a maximally entangled Bell state to reach $p_{\text{guess}}=1$, certifying no secure randomness. Expanding the POVM to include non-orthogonal bases introduces non-zero correlations between POVM elements, limiting Eve’s ability to predict measurement outcomes and thereby ensuring consistent randomness across all input polarizations. From the graph-theoretic perspective discussed, these correlations manifest as non-zero edge weights, resulting in non-zero weighted degrees for vertices, which signify robust, input-polarization-independent certified randomness. Greater edge weights correspond to stronger element correlations, leading to increased randomness and a reduced guessing probability for Eve. If an isolated vertex with zero weighted degree exists, one again has $p_{\text{guess}}=1$, whereas an optimal, symmetric configuration in which all vertices have equal weighted degree minimizes $D_{i}$ and thus maximizes the certified quantum randomness.
	
	The certified randomness per measurement response, constrained by the weighted degrees of the POVM graph, can be quantified by the conditional min-entropy $ H_{\text{min}}(A | E) $, expressed as
	\begin{eqnarray}
		H_{\text{min}}(A|E)&=&-\log_{2}\left (g_{m} \: p_{\text{guess}}(A|\rho_{A}^{m}, \{F_{i}\}_{N})\right )\nonumber\\&\geq&m\log_{2}\left (\min_{i}D_{i}\right )-\log_{2}g_{m} \label{hm}
	\end{eqnarray} 
	Since photons are bosons, detection events are indistinguishable, ignoring their order, and the response counts correspond to the photon number $ m $. Thus, the coefficient $ g_m $ is defined as
	\begin{eqnarray}
		g_{m}=\dfrac{\sum_{i1,i2,...,im}\eta_{i1}\eta_{i2}\cdots \eta_{im}}{\sum_{i1<i2<...<im}\eta_{i1}\eta_{i2}\cdots \eta_{im}} \qquad i1,i2,...,im\in\{1,2...N\}
	\end{eqnarray}
	
	Beyond the single-round scenario, our results also ensure composable security against arbitrary coherent attacks, where the input states in different rounds may be entangled into a multi-round state (see Supplementary Materials S4 \cite{SupplementalMaterial}). In Fig. \ref{fig:2}(c), we present the conditional min-entropy for a metasurface with $N$ output ports uniformly spaced in linear polarization angle across the Bloch plane, as a function of $N$ and the input photon number $m$. As $N$ increases, the min-entropy rises monotonically, which is driven by the emergence of more weighted edges in the POVM graph, as each new port adds a vertex and introduces non-zero correlations via non-orthogonal Jones matrices. For a fixed $N$, the min-entropy reaches its maximum when $m \approx N/2$, which is intuitive: this regime maximizes the number of distinct detection events, balancing channel occupancy and outcome distinguishability. 
	
	For realistic light sources exhibiting mixed Fock-state statistics, such as Poisson or thermal distributions which are shown in Fig. \ref{fig:2}(d), the total certified randomness must account for the photon-number distribution under the worst-case eavesdropping attack. The overall min-entropy is thus (see Materials and Methods)
	
	\begin{eqnarray}
		&&H_{\text{min}}(A_{\tau}|E)\geq \sum_{m=1}^{N-1}\gamma_{m}m\log_{2}\left (\min_{i}D_{i}\right )-\log_{2}g_{m}\\
		&&\gamma_{m}=\frac{\tau_{m}\delta^{m}}{\tau_{m}\delta^{m}+\sum_{i> m}\tau_{i}}Q_{m} \label{hmc}
	\end{eqnarray}
	where $\tau_m$ is the probability of emitting $m$ photons in the source, $Q_m$ is the probability of registering $m$ detection clicks, and $\delta$ is the overall detection efficiency. In Fig. \ref{fig:2}(d), we show the min-entropy for a Poisson-distributed source as a function of channel number $N$ and mean photon number $\mu$. Unlike fixed-$m$-photon inputs, Poisson-distributed inputs suffer increased entropy loss at high mean intensity due to greater exposure to eavesdropping attacks, while an excessive number of channels dilutes the per-port response probability.

	\subsection*{\label{sec:level3}Experiment}
	
	To experimentally realize the proposed POVM graph and achieve polarization-independent quantum randomness generation, we designed and fabricated metasurfaces that support $N = 4$ and $N = 6$ output channels. TiO$_2$ was used as the pillar material with a thickness of 1000 nm, deposited on a dry-cleaned fused silica substrate, as shown in Fig~\ref{fig:3}(a) (see Supplementary Materials S1-S2 \cite{SupplementalMaterial}). The $N = 4$ metasurface, designed using a matrix Fourier optics approach, supports four primary diffraction orders, including two sets of orthogonal linear polarization bases. The $N = 6$ metasurface extends this to six diffraction orders by incorporating an additional set of orthogonal polarization bases, enabling richer POVM structures and potentially higher randomness generation.

	To validate the functionality of the metasurfaces, we performed classical characterization by measuring the intensity ratios at the output channels using an 808 nm laser. Various input polarization states were generated by rotating a half-wave plate (HWP) or a quarter-wave plate (QWP). The output light from each channel was collected via fiber couplers and detected with photodiodes. The measured results, shown as dots in Fig.~\ref{fig:3}(d) for the six-channel case, agree well with the theoretical predictions lines. From these measurements, we reconstructed the Jones matrix for each diffraction channel. In Fig.~\ref{fig:3}(b, c), we present the corresponding Jones vectors visualized on the Bloch sphere for both the $N = 4$ and $N = 6$ metasurfaces, together with their quantum POVM graphs. These dual representations clearly illustrate the distribution and near-orthogonality of the realized polarization projection bases, where opposite vertices have almost no connecting edges. The weighted degree of each vertex, computed using Eq.~\ref{de}, is explicitly indicated in the following, directly linking the geometric configuration on the Bloch sphere to the strength of correlations between POVM elements. Imperfections in functionality and nonuniform channel efficiencies manifest as unequal vertex weighted degrees.
	
	Next, we demonstrate single-photon randomness generation using the fabricated metasurfaces, as shown in Fig.~\ref{fig:4}(a). A heralded single-photon source is generated via spontaneous parametric down-conversion (SPDC) in a type-I BBO crystal. The photon pairs are separated, and the signal photon is focused onto the metasurface using a 4f beam-reducing system. Detection is performed with a single-photon detector (SPD) coupled to a time-division multiplexing (TDM) system. In this setup, 500 ns time windows are assigned to different diffraction channels, enabling sequential detection of photons from all output ports using a single SPD via fiber coupler. Coincidence events are recorded through correlation analysis between the herald and signal photons, with precise time tagging used to assign each detection to its corresponding diffraction channel.
	
	In Fig.~\ref{fig:4}(b--d), we compare the conditional min-entropy of our fabricated 4- and 6-channel metasurfaces obtained from our graph-theoretic bound with that obtained using the traditional QRNG estimation (see Supplementary Materials S6 \cite{SupplementalMaterial}), which directly relates the adversary’s guessing probability to the maximum output probability. For pure linearly polarized input states, the traditional estimation exhibits strong state dependence, whereas our method yields polarization-independent bounds that remain secure for any input polarization, as shown in Fig.~\ref{fig:4}(b). For mixed input states, however, the situation is fundamentally different: under the traditional estimation, the randomness can become completely insecure. An adversary who is correlated with the mixed states source can fully exploit this by controlling the polarization decomposition and the ordering of the polarization components, or by preset entanglement with part of our system. Prior knowledge of the specific decomposition (or, equivalently, preset entanglement, see Supplementary Materials S7 \cite{SupplementalMaterial} for details) allows the adversary to achieve a much higher guessing probability than suggested by the traditional estimation, while producing output statistics that are indistinguishable for the attack at the level of observed probabilities. As illustrated by the progression from Fig.~\ref{fig:4}(c) to (d), the more mixed the input state, the larger this unsecure gap becomes. In contrast, our input-independent bound remains valid and secure for arbitrary input states and against any such attack strategy.

	We also implemented randomness generation using a mixed Fock-state source generated by a 300 kHz pulsed laser, producing a Poisson photon number distribution. A variable optical attenuator was inserted to precisely control the mean photon number to the single-photon regime. As shown in Fig.~\ref{fig:4}(e), the optimal average photon number is approximately 0.4 photons per pulse, balancing randomness yield and multi-photon noise. Although mixed Fock states introduce additional multi-photon components that reduce the conditional min-entropy due to enhanced eavesdropping vulnerability in security analysis, this approach offers significant practical advantages: simplified setup, high repetition rate, and ease of integration. These benefits strongly motivate its use in scalable QRNG systems.
	
	The detection outcomes from different diffraction channels were encoded using 2-bit (4-channel) and 3-bit (6-channel) labels to generate raw bit sequences. Final secure randomness was extracted via a Toeplitz-matrix-based extractor, with the output length determined by the conditional min-entropy and extractable randomness bounds, as discussed in Method. With the heralded single-photon source, the $N=4$ metasurface yields 0.6153 bits of certified randomness per detection, while the $N=6$ metasurface achieves 0.8059 bits per detection. Using the pulsed laser (Poisson-distributed) source, the $N=4$ and $N=6$ metasurfaces generate 0.1192 bits and 0.1220 bits of randomness per response, respectively. The extracted bit streams were subjected to the NIST statistical test suite \cite{rukhin2001statistical}, successfully passing all tests and confirming excellent statistical randomness properties (see Supplementary Materials S13 \cite{SupplementalMaterial}).

	\section*{Discussion and outlook}
	
	Our work establishes a graph-theoretic paradigm for quantum randomness generation using resonant nanophotonic metasurfaces, enabling polarization-independent certified randomness through parallel, high-dimensional POVM measurements within a single ultrathin layer. By mapping metasurface diffraction channels onto graph vertices and non-orthogonal Jones-matrix overlaps onto weighted edges, the resulting POVM graph directly connects metasurface design and graph topology to weighted degrees that determine the conditional min-entropy, thereby allowing the randomness performance of a device to be predicted and optimized even when the input state is arbitrary, mixed, or entangled with an adversary under the adopted security model. 
	
	Unlike conventional QRNG analyses that infer security primarily from observed output probabilities and may therefore be vulnerable to source correlations, our approach derives input-state-independent guarantees from the complete POVM structure. It also advances graph theory in metasurface quantum optics beyond its previous descriptive role in visualizing channel-dependent bunching or antibunching in multichannel HOM interference \cite{yousef2025metasurface}, transforming it into a quantitative tool for entropy evaluation and semi-device-independent security certification under realistic imperfections. From an implementation perspective, our metasurface platform realizes high-dimensional POVMs and their security analysis in a compact, passive, and alignment-free architecture, compressing complex multi-outcome measurements into a single ultrathin element suitable for scalable QRNGs. Because the framework connects general graph structures to entropy bounds, it may also be extended to other high-dimensional photonic platforms, advanced QRNG protocols \cite{herrero2017quantum}, measurement certification \cite{coles2017entropic}, and high-dimensional quantum key distribution \cite{xu2020secure}.

	\section*{\label{sec:leveled}Materials and Methods}

	\subsection*{Guessing probability by weighted degree}
	
	Randomness estimation hinges on the chosen security level. As detailed above, we consider an unknown source emitting $m$-photon states $\rho_A^m$, subject to an entanglement-based attack where Eve holds the purification $|\psi_{AE}\rangle$ and performs an optimal POVM $\{M_i^E\}_N$ on her subsystem to guess the measurement outcomes. The metasurface implements an $N$-channel POVM $\{F_i\}_{i=1}^N$, and the guessing probability is:
	\begin{eqnarray}
		p_{\text{guess}}(A|\rho_{A}^{m}, \{F_{i}\}_{N})&=&\max_{\{\{M^{E}_{i}\}_{N},\ket{\psi_{AE}}\}}\sum_{i}\Tr[F_{i}^{\otimes m}\otimes M^{E}_{i}\ketbra{\psi_{AE}}]\nonumber\\&=&\max_{\{p_{i},\sigma_{Ai}\}}\sum_{i}p_{i}\Tr[F_{i}^{\otimes m}\sigma_{Ai}]
	\end{eqnarray}
	where $\sigma_{Ai}$ is the reduced state after tracing out Eve
	\begin{eqnarray}
		\sigma_{Ai}=\frac{\Tr_{E}[I^{\otimes m}\otimes M^{E}_{i}\ketbra{\psi_{AE}}]}{\Tr[I^{\otimes m}\otimes M^{E}_{i}\ketbra{\psi_{AE}}]}
	\end{eqnarray}
	normalized such that $\sum_{i}p_{i}=1$ with $p_{i}=\Tr[I^{\otimes m}\otimes M^{E}_{i}\ketbra{\psi_{AE}}]$. From the POVM definition, the probability of outcome $i$ for input $\sigma_{A i}$ is:
	\begin{eqnarray}
		\Tr(F_{i}^{\otimes m} \sigma_{Ai})&=&\begin{aligned}[t]\Tr_{1}(F_{i}\otimes I_{2}\otimes\cdots\otimes I_{m} \,&\Tr_{2}(I_{1}\otimes F_{i}\otimes\cdots\otimes I_{m}\Tr_{3}\cdots \\&\Tr_{m}(I_{1}\otimes\cdots\otimes I_{m-1} \otimes F_{i} \sigma_{Ai})\cdots))\end{aligned}\nonumber\\
		&=&	\begin{aligned}[t]
			\Tr_{1}(t \eta_i J_i\otimes I_{2}\otimes\cdots\otimes I_{m} \,&\Tr_{2}(I_{1}\otimes t \eta_i J_i\otimes\cdots\otimes I_{m}\Tr_{3}\cdots \\&\Tr_{m}(I_{1}\otimes\cdots\otimes I_{m-1} \otimes t \eta_i J_i \sigma_{Ai})\cdots))
		\end{aligned}
	\end{eqnarray}
	Since $J_i \leq I$ (which means that \(I-J_i\) is positive semidefinite) and $\sigma_{A i}$ is a valid quantum state, we have the physical bound
	\begin{eqnarray}
		\Tr(F_{i}^{\otimes m} \sigma_{Ai})\leq	(t\eta_{i})^{m}
	\end{eqnarray}
	To derive an upper bound on $t \eta_i$, we exploit the POVM completeness relation $\sum_{i=1}^N t \eta_i J_i \le I,$ which implies that the sum of all scaled Jones operators cannot exceed the identity in the Loewner order. Taking the Hilbert–Schmidt inner product with $J_j$ on both sides yields
	\begin{eqnarray}
		t\sum_{i=1}^{N} \eta_{i} \langle J_{i},J_{j} \rangle  \leq \langle J_{i},I \rangle\leq1
	\end{eqnarray}
	Remarkably, this bound directly connects the normalization parameter $t$ to the total weighted edge strength in the POVM graph. Using the weighted degree of vertex $i$, defined as $	D_{i}= \frac{1}{\eta_{i}}\sum_{j=1}^{N} \eta_{j} \langle J_{i},J_{j} \rangle  $, for the $i$ outcome probability under the input state $\sigma_{Ai}$, we have
	\begin{eqnarray}
		\Tr(F_{i}^{\otimes m} \sigma_{Ai})\leq	D_{i}^{-m}
	\end{eqnarray}
	These individual outcome probabilities directly bound Eve’s optimal guessing probability for the effective $m$-photon detection event
	\begin{eqnarray}
		p_{\text{guess}}(A|\rho_{A}^{m}, \{F_{i}\}_{N})\leq\max_{\{p_{i}\}}\sum_{i}p_{i}D_{i}^{-m}\leq\max_{i}\{D_{i}^{-m}\}
	\end{eqnarray}
	Because photons are indistinguishable bosons, detection events are order-insensitive, and coincident photons in the same output port cannot be distinguished from losses (see Fig.~\ref{fig:g} for the $m=2$ case). Therefore, only events in which exactly $m$ distinct ports fire are considered valid $m$-photon responses. This restriction reduces the number of observable outcomes and is captured by the coefficient
	\begin{eqnarray}
		g_{m}=\dfrac{\sum_{i1,i2,...,im}\eta_{i1}\eta_{i2}\cdots \eta_{im}}{\sum_{i1<i2<...<im}\eta_{i1}\eta_{i2}\cdots \eta_{im}}\qquad i1,i2,...,im\in\{1,2...N\}
	\end{eqnarray}
	where $\eta_{i1},\eta_{i2},... ,\eta_{im}$ is the transmission efficiency of port $i1,i2,...,im$. A larger $g_m$ corresponds to fewer distinguishable multi-photon events, which increases Eve’s maximum guessing probability and consequently reduces the certified randomness per detected $m$-photon event.

	\subsection*{Condition min-entropy estimation}
	
	The certified randomness per measurement response is quantified by the conditional min-entropy $H_{\min}(A|E)$. Following standard cryptographic security definitions \cite{konig2009operational}, we adopt the min-entropy conditioned on Eve’s quantum side information $E$, formally defined as 
	
	\begin{eqnarray}
		H_{\text{min}}(A|E)&=&-\inf_{\sigma_{E}}\{\lambda\in \textbf{R}: \rho_{AE}\leq 2^{\lambda}\omega_{A}\otimes\sigma_{E}\}\nonumber\\&=&-\log_{2} p_{\text{guess}}(A|E)
	\end{eqnarray} 
	where the infimum ranges over all normalized density operators $\sigma_E$, and $\omega_A$ is the fully mixed state. $\rho_{AE}$ is the joint state of the measurement outcome $A$ and Eve’s system $E$. From the guessing probability bound derived in our graph-theoretic framework, the conditional min-entropy for an $m$-photon event is rigorously lower-bounded by
	\begin{eqnarray}
		H_{\text{min}}(A|E)&=&-\log_{2}\left (g_{m} \: p_{\text{guess}}(A|\rho_{A}^{m}, \{F_{i}\}_{N})\right )\nonumber\\&\geq&m\log_{2}\left (\min_{i}D_{i}\right )-\log_{2}g_{m} 
	\end{eqnarray} 
	This bound directly translates the weighted-degree structure of the POVM graph into a quantifiable, composable security guarantee for the extracted random bits, even in the presence of arbitrary eavesdropping attacks.
	
	\subsection*{Mixed Fock-state source}
	
	Realistic light sources typically exhibit mixed Fock-state statistics, such as Poisson or thermal distributions, which are easier to implement than heralded single- or fixed-number multi-photon sources. When the input contains a mixture of different photon numbers, higher-photon-number components can bunch into the same output ports, masking the responses of lower-photon-number events and potentially compromising security. To achieve a rigorous, composable security proof, we adopt the worst-case eavesdropping model, in which any detection event registering exactly $m$ clicks is conservatively assumed to originate only from input states with $m$ photons (higher-photon-number components are fully controlled by Eve \cite{wang2005beating}). This leads to an effective randomness efficiency for $m$-click events of
	\begin{eqnarray}
		\gamma_{m}&=&\frac{\tau_{m}\delta^{m}}{\sum_{i\geq m}\tau_{i}\binom{i}{m}\delta^{m}(1-\delta)^{i-m}}Q_{m}\nonumber\\
		&\geq&\frac{\tau_{m}\delta^{m}}{\tau_{m}\delta^{m}+\sum_{i> m}\tau_{i}}Q_{m}
	\end{eqnarray}
	where $\tau_m$ is the probability of emitting $m$ photons in the source, $Q_m$ is the probability of registering $m$ detection clicks, and $\delta$ is the overall detection efficiency. The overall conditional min-entropy, secure against worst-case attacks, is then lower-bounded by
	\begin{eqnarray}
		&&H_{\text{min}}(A_{\tau}|E)\geq \sum_{m=1}^{N-1}\gamma_{m}m\log_{2}\left (\min_{i}D_{i}\right )-\log_{2}g_{m}
	\end{eqnarray}
	This expression provides a tight, experimentally achievable bound on the final extractable randomness rate for practical, uncharacterized light sources.

	\clearpage

	\begin{figure}[htbp]
		\centering
		\includegraphics[width=6.5in]{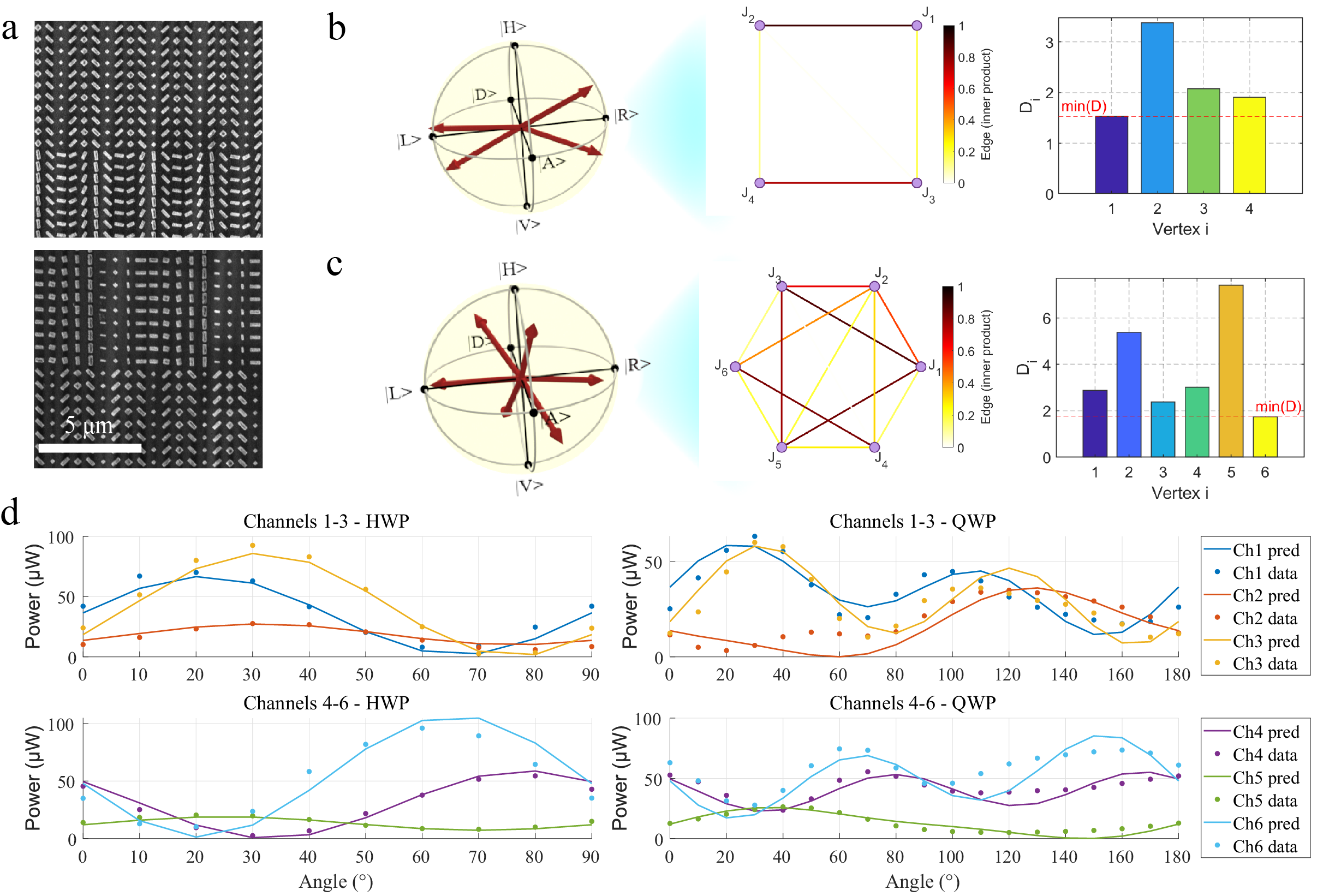}
		\caption{Characterization of multi-channel TiO$_{2}$ metasurfaces. (a) Scanning electron microscopy images of the fabricated TiO$_{2}$ 6-channel metasurfaces (scale bars: 5 $ \mu$m). (b, c) POVM Bloch vectors associated with the diffraction orders of the practical 4 and 6-channel metasurfaces (left), the corresponding weighted quantum graphs (middle), and the weighted degree of each vertex (right). The minimum-degree vertices correspond to $J_{1} $ in the 4-channel case and $J_{6} $ in the 6-channel case. (d) Measured output power in each diffraction order of the 6-channel metasurface as a function of the angle of a half-wave plate (HWP) or quarter-wave plate (QWP) placed before the device to prepare different input polarization states. Experimental data are shown as dots; solid lines are theoretical predictions based on the experimentally calibrated transfer matrix of the metasurface. }
		\label{fig:3}
	\end{figure}
	
	\begin{figure*}[htbp]
		\centering
		\includegraphics[width=6.5in]{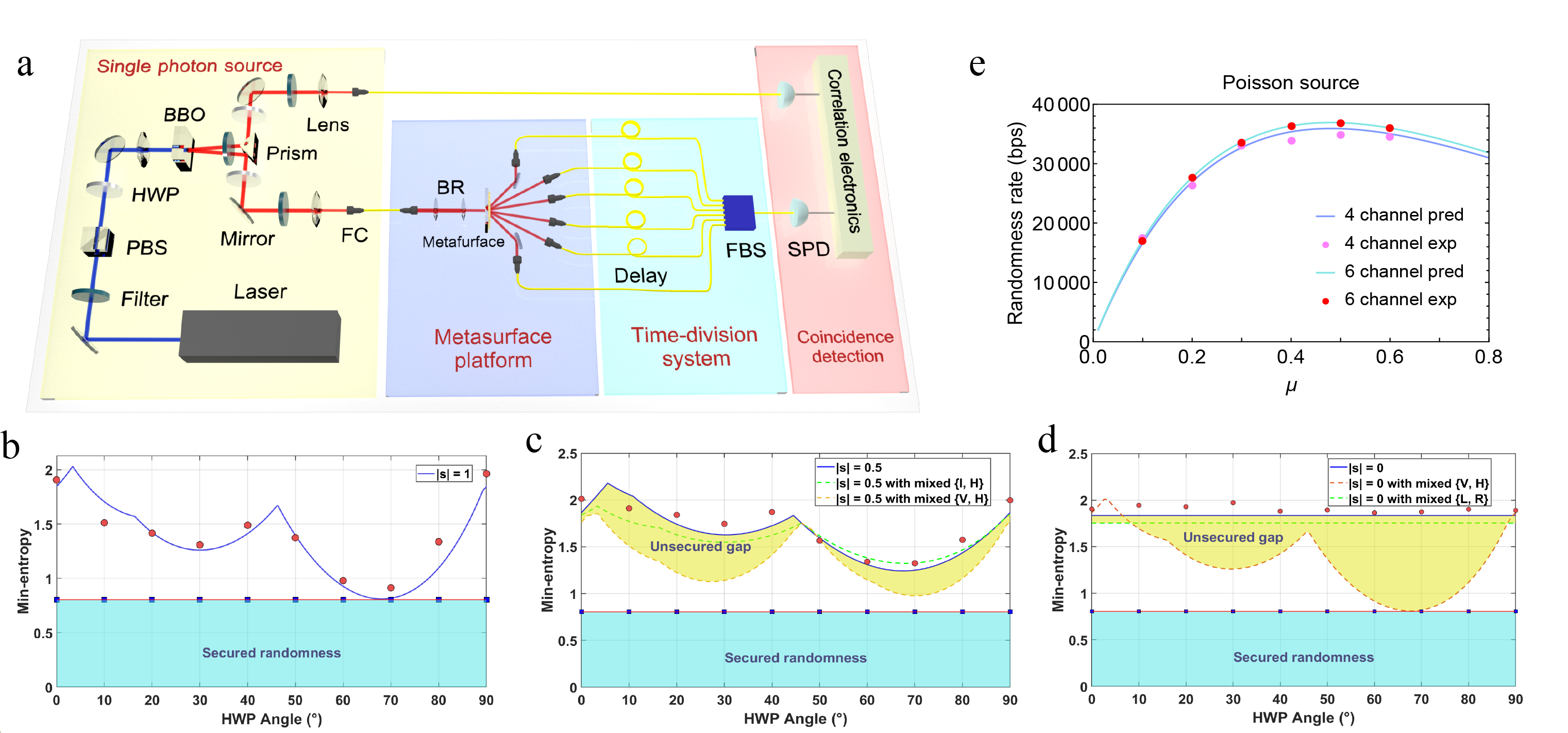}
		\caption{Experiment results for our QRNG based on metasurface quantum graph. (a) Schematic of the experimental setup: a heralded single-photon source generated via type-1 SPDC in a 0.5 mm BBO crystal, followed by the metasurface test platform, temporal multiplexing via fiber delay lines, and coincidence counting electronics. PBS, polarization beam splitter; HWP, half-wave plate; FC, fiber coupler; BR, beam reducer; FBS, fiber beam splitter; SPD, single-photon detector. (b--d) Min-entropy as a function of the input state for the 6‑channel metasurfaces, comparing the graph-theoretic bound (light green shaded region) with the randomness inferred directly from output probabilities (blue line: simulation; red points: experiment). (b): pure H-polarized input with varying HWP angle, showing strong state dependence of probability-based randomness in contrast to the polarization-independent graph-theoretic bound. (c, d): mixed polarization inputs with Bloch-vector modulus $\abs{s}=0.5 $ (c) and $\abs{s}=0 $ (d), scanned via HWP angle; probability-based randomness overestimates the secure randomness because an adversary correlated with the chosen decomposition (green and orange dashed lines) can gain extra information, whereas the graph-theoretic bound remains secure and state independent. (e) Randomness generation rate for a coherent source input with Poissonian photon-number statistics for the 4- and 6-channel metasurfaces.}
		\label{fig:4}
	\end{figure*}
	
	\begin{figure}[htbp]
		\centering
		\includegraphics[width=3.5in]{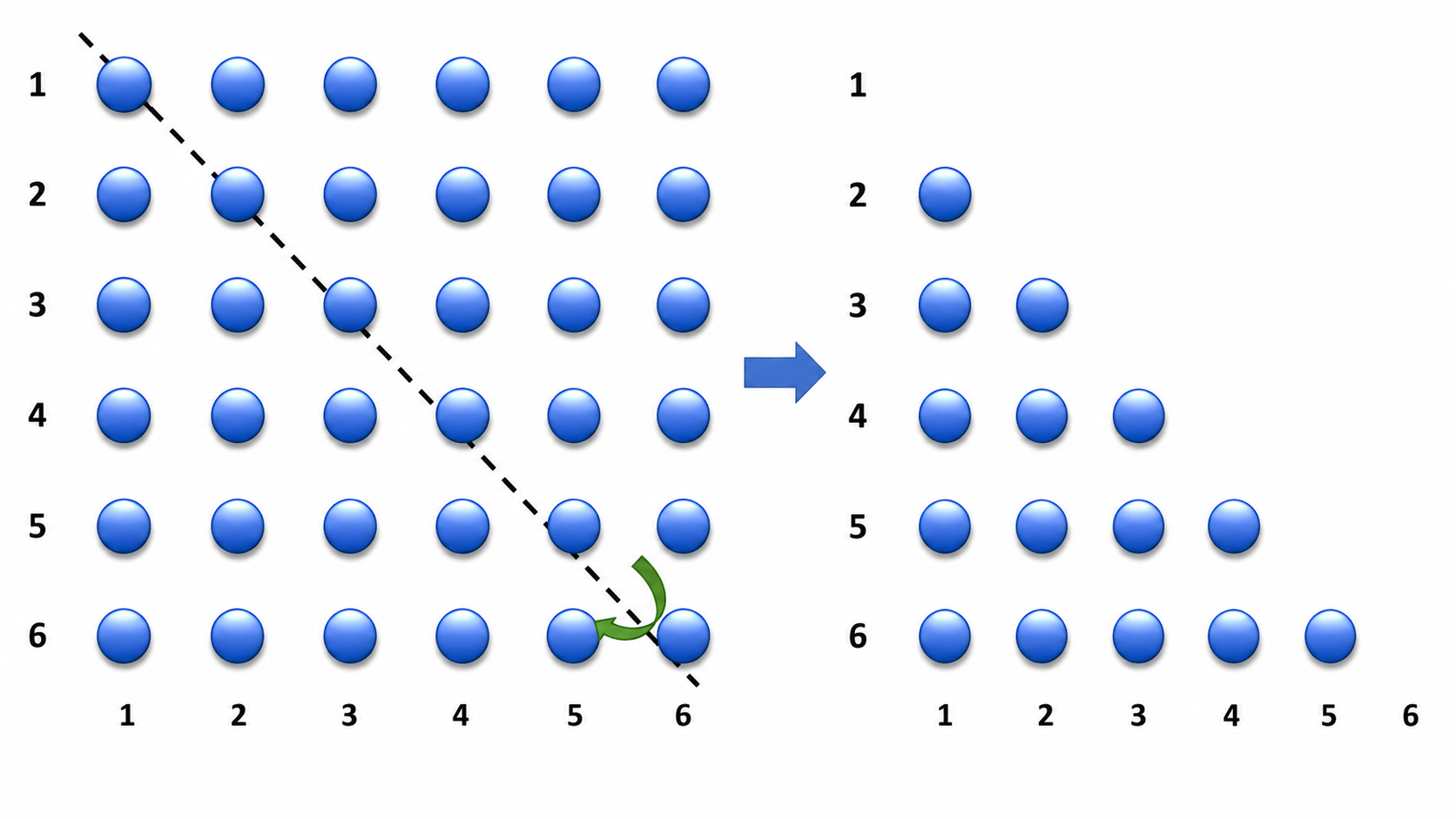}
		\caption{Schematic illustration of valid and invalid detection events for the $m=2$ case. Because photons are indistinguishable bosons and the detectors are insensitive to photon number, only runs in which two distinct output ports fire are accepted as valid two-photon responses, and this post-selection generalizes to requiring exactly $m$ distinct ports for an $m$-photon input.}
		\label{fig:g}
	\end{figure}

	\clearpage
	
	\nocite{*}
	\bibliographystyle{ieeetr}
	\bibliography{article}

	\section*{Acknowledgments }
	
	We thank Yuri Kivshar, Ruixuan Zheng, Qinyang Mo and Ye Tian for useful discussions and assistance. 
	
	\paragraph*{Funding: }
	The work at The University of Hong Kong was sponsored by the New Cornerstone Science Foundation, the Research Grants Council of Hong Kong (AoE/P-502/20, STG3/E-704/23-N, 17309021), the Guangdong Provincial Quantum Science Strategic Initiative (GDZX2204004, GDZX2304001). National Natural Science Foundation of China (Nos. 12522421), National Key R\&D Program of China (No. 2023YFA1407700)
	
	\paragraph*{Author contributions}
	Conceptualization: X.L. and S.Z. Theory analysis and simulations: X.L. Samples simulations: W.W.L. Fabricated samples: W.G.S. Quantum source building: X.L. and S.X.C. Test experiment: X.L. Data analysis and interpretation: X.L. and S.Z. Writing—original draft, review and editing: X.L., W.W.L. and S.Z. Supervision: S.Z.

	\paragraph*{Competing interests:}
	
	The authors declare that they have no competing interests.

	\paragraph*{Data, code and materials availability: }
	
	All data and code needed to evaluate and reproduce the results in the paper are present in the paper and/or the Supplementary Materials. 
	
	\subsection*{Supplementary materials}
	
	Supplementary Texts S1 to S13
	
	\noindent Figures S1 to S7
	
	\noindent References

	\clearpage
	
\setcounter{equation}{0}
\setcounter{figure}{0}
\setcounter{subsection}{0}
\renewcommand{\theequation}{S.\arabic{equation}}
\renewcommand{\thefigure}{S\arabic{figure}}
\renewcommand{\thesubsection}{S\arabic{subsection}}

\begin{center}
	% 论文题目（与主文一致）
	{\Large Supplementary Materials for \\[0.5em]\textbf{Metasurface quantum graphs for polarization-independent certified randomness }}\\[1.5em]
	
	% 作者（示例：Thomas Jones et al.）
	{\normalsize Xing Lin, et al}\\[0.5em]
	
	% 通讯作者
	{\footnotesize Corresponding author: Shuang Zhang, shuzhang@hku.hk}\\[2em]
\end{center}

\noindent\textbf{This PDF file includes:}\\

Supplementary Texts S1 to S13

Figures S1 to S7

References

\newpage

\subsection{Material design principle}	

The metasurface consists of numbers of interleaved metagratings, each designed to diffract a specific pair of elliptical polarization states into two directions symmetric about the transmission axis. Such geometric-phase metagrating can arbitrarily select and project any desired pair of elliptical polarization states with great flexibility, which is in contrast with classical polarization analysis that reconstructing multi-photon quantum states with on-off (click) detectors requires access to multiple elliptical polarization components. 

Here, we use the combination of geometric (Pancharatnam–Berry) and propagation phases [34], which enables independent imposition of arbitrary phase profiles on any pair of orthogonal polarization states—whether linear, circular, or elliptical. 

\begin{figure*}[htbp]
	\centering
	\includegraphics[width=6.5in]{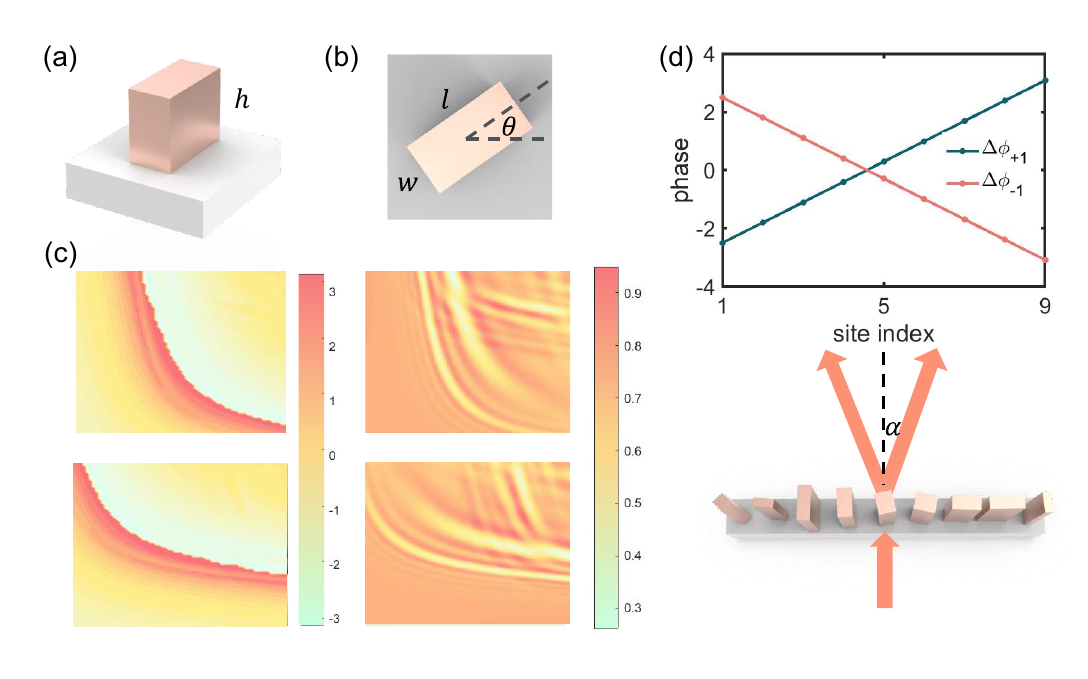}
	\caption{(a-b) The perspective view (a) and top view (b) of the single atom design of the birefringent meta-atom. The hight we use equals to 1000nm while other parameters including the length (l) width (w) and rotation angle ($\theta$) are vary. (c) The amplitude and phase distribution of the single atom with varying length and width. (d) The spatial phase distribution when light is going through the super-atom surface that made with 9 different atoms shown below. The phases show a gradient distribution that contribute to the radiation angle of $\alpha$.}
	\label{fig:metad}
\end{figure*}

The building block of the polarization splitting material is a rectangular meta-atom with height $h$ that behaves as a birefringent crystal as shown in Fig. \ref{fig:metad}(a-b). The corresponding Jones matrix is 

\begin{eqnarray}
	J^\prime=R(\theta)\left[\begin{matrix}e^{i\delta_f}&0\\0&e^{i\delta_s}\\\end{matrix}\right]R(-\theta)
\end{eqnarray}
in which $R(\theta)$ is the rotation matrix that equals to $\left[\begin{matrix}\cos{\theta}&-\sin{\theta}\\\sin{\theta}&\cos{\theta}\\\end{matrix}\right]$. $ \delta_f$ and $\delta_s$ is the phase shift on light linearly polarized along the fast and slow axes, which can be individually tailored by adjusting its shape, giving rise to the propagation phase. The difference between the phase shift determines the polarization conversion phase. Meanwhile, the angular orientation $\theta$ (Fig. \ref{fig:metad}(b)) could give a phase profile on the state that equals to $2\theta$ varying over spatial extent, which we call geometry phase. Such spatial orientation allows the polarization transformation with desired output angle. 

From above discussion, we could control both the geometric phase and propagation phase by varying the shape and orientation of each birefringent meta-atom. In the following, we show that arbitrary desired polarization state pairs scattering into two directions with respect to the transmission axis can be formed. Suppose the following desired Jones matrix:
\begin{eqnarray}
	J\left(x,y\right)=\left[\begin{matrix}e^{i\phi^+\left(x,y\right)}\kappa_1^+&e^{i\phi^-\left(x,y\right)}\kappa_1^-\\e^{i\phi^+\left(x,y\right)}\kappa_2^+&e^{i\phi^-\left(x,y\right)}\kappa_2^-\\\end{matrix}\right]\left[\begin{matrix}\lambda_1^+&\lambda_1^-\\\lambda_2^+&\lambda_2^-\\\end{matrix}\right]^{-1}
\end{eqnarray}
where $\phi^+(x,y)$ and $\phi^-(x,y)$ represent the arbitrary spatial phase profiles on the states and $\lambda^\pm=\begin{pmatrix} \lambda^\pm_{1} \\ \lambda^\pm_{2}	\end{pmatrix}$ are the two extracted orthogonal input orthogonal polarization states while $ \kappa^\pm=\begin{pmatrix} \kappa^\pm_{1} \\ \kappa^\pm_{2}	\end{pmatrix}$ are the output orthogonal states. The input and output states are linked by opposite handedness. The target bases are chosen from the Bloch sphere depending on orthogonality among bases as we will discuss in the Section S3. The $\phi^\pm(x,y)$ is decided by the desired radiation angle: 
\begin{eqnarray}
	\mathrm{\Theta}_{output}=\arcsin{(\frac{\lambda\Delta\phi}{2\pi\Delta x})}
\end{eqnarray}
By making $J^\prime=J$, we could calculate $\theta$, $\delta_f$ and $\delta_s$ for each meta-atom. By laying meta-atoms that provide gradient spatial phase $\Delta\phi$ with distance $\Delta x$ for the same polarization state, one could get the super-atom that generates arbitrary polarization state pairs in two directions.

\subsection{Meta-grating arrangement and experiment}	

\begin{figure*}[htbp]
	\centering
	\includegraphics[width=6.5in]{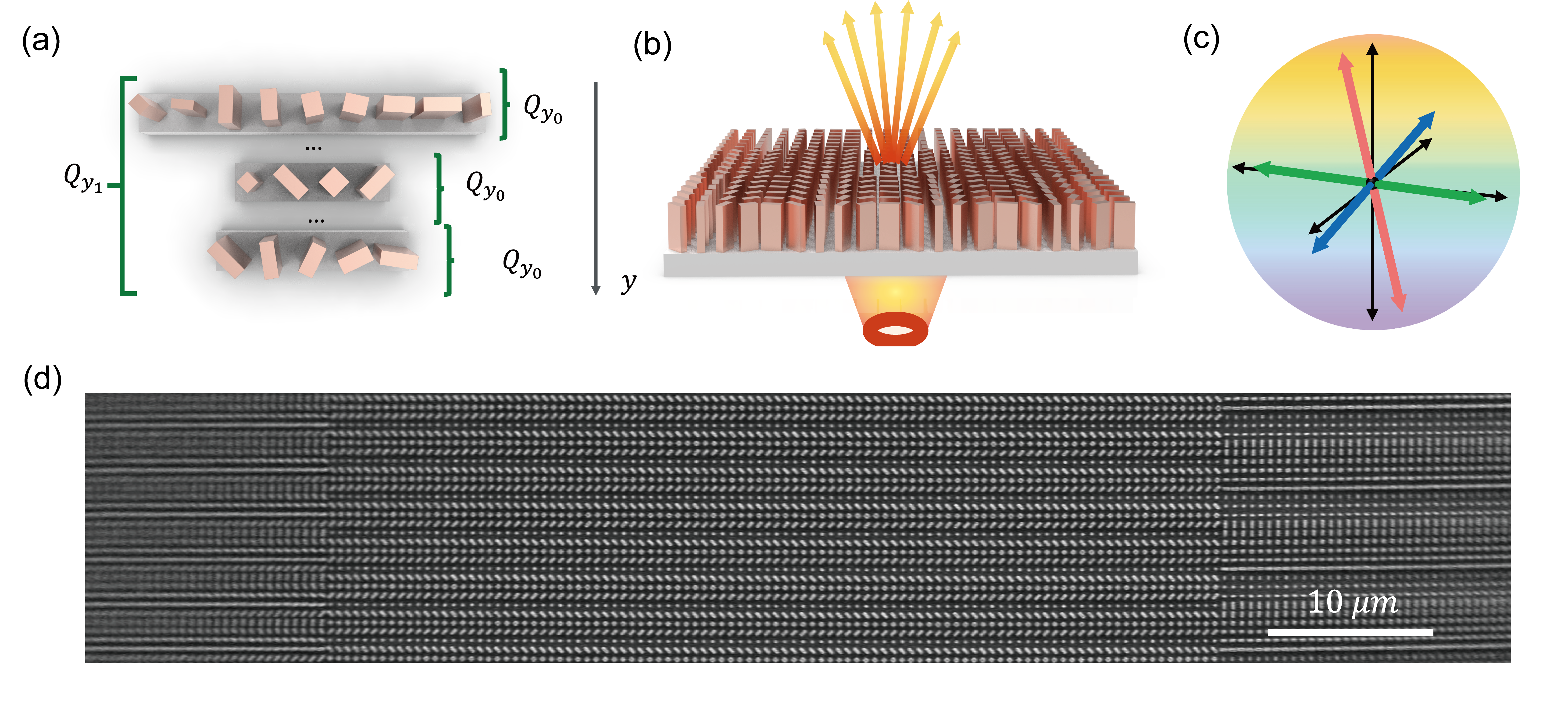}
	\caption{(a) Illustration of spatial distribution of three set of different super atoms. Each super atom repeats periodically along x direction and repeat 100 times followed by the next kind in the y direction. (b) Schematic view of the metasurface that interacts with photons which has six output channels. (c) The corresponding six polarization states of the three-component structure. (d) The SEM of the metasurface.}
	\label{fig:metad2}
\end{figure*}

In our work, the target is to provide 4 and 6 polarization states using one metasurface. Here, we use six states as example to illustrate the arrangement method. To have those six states, we need to design three kinds of super-atoms each providing a pair of orthogonal polarization state. Fig. \ref{fig:metad2}(a) show the super-atom design and spatial phase distribution for each kind, corresponding to different polarization conversion and scattering angle. Then each super-atom repeats $N_i=P/a/n_i (i=1,2,3)$ times along the grating direction x while super-atoms repeat $M_i=100$ times followed by the next series $[M_1,M_2, M_3]$ in the y direction to generate six polarization states as also shown in Fig. \ref{fig:metad2}(a). The number of cells is determined following the rule [34] to enhance diffraction efficiency and minimize the dependence of the transfer matrix on the spatial beam profile. The whole metasurface size equals to 2 mm$\times$2 mm. When photons with arbitrary polarization state are injected into the metasurface, they will split into six output channels with different angles as illustrate in Fig. \ref{fig:metad}(b). The corresponding polarization of these channels are shown in Fig. \ref{fig:metad}(c).                            

\begin{figure*}[htbp]
	\centering
	\includegraphics[width=6.5in]{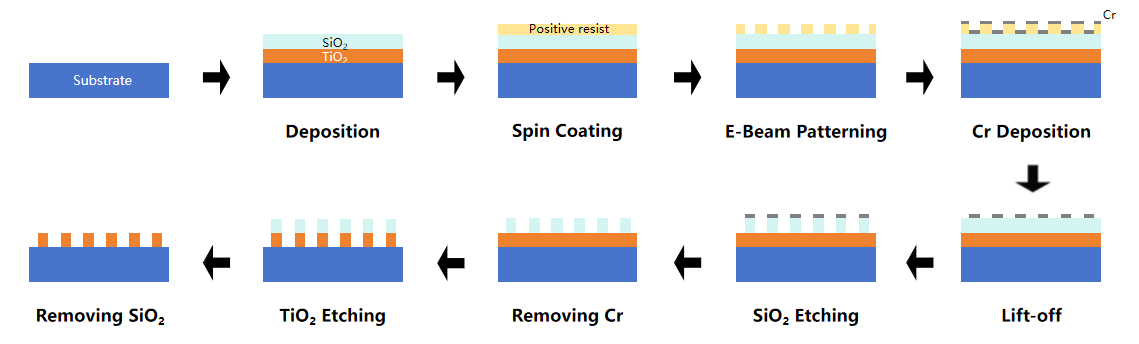}
	\caption{Process flow for fabricating TiO$_2 $ nanostructures. Starting from a bare substrate, TiO$_2 $ and SiO$_2 $ layers are deposited, followed by spin coating of a positive resist and e‑beam patterning. After Cr deposition and lift‑off to form the Cr hard mask, the pattern is transferred by sequential SiO$_2 $ etching, Cr removal, TiO$_2 $ etching, and final removal of the remaining SiO$_2 $ layer.}
	\label{fig:pro}
\end{figure*}

Fig. \ref{fig:metad2}(d) shows the Scanning Electron Microscope (SEM) images of our periodic grating metasurface. In experiment, we choose TiO$_{2}$ as material of the pillars. Our central working wavelength is 808 nm. The TiO$_{2}$ layer was deposited on the dry-cleaned fused silica substrate to a final thickness of 1000 nm. Then a layer of photoresist film was spin-coated onto the substrate and baked. A layer of water-soluble conductive polymer was spin coated on the resist for the dissipation of E-beam charges. The device pattern was written on the electron beam resist using an E-beam writer. An electron beam evaporated chromium layer was used to reverse the generated pattern with a lift-off process and was then used as a hard mask for dry etching the TiO$_{2}$ layer. After that, the devices transferred into the etcher. Finally, the chromium layer was removed by a stripping solution (ceric ammonium nitrate), as shown in Fig. \ref{fig:pro}. 

\subsection{\label{sec:level51}POVM graph representation}

In the POVM graph, each vertex corresponds to a diffraction channel characterized by its Jones matrix $J_i$, which encodes the polarization transformation of the associated optical path. Pure polarization states are not required; the Jones matrix may represent mixed or partially polarized light, meaning the corresponding polarization vector lies inside the Bloch sphere, not necessarily on its surface. The correlation between any two vertices $i$ and $j$ is quantified via the Hilbert–Schmidt inner product of their Jones matrices:
\begin{equation}
	\langle J_i, J_j \rangle = \frac{\Tr(J_i^\dagger J_j)}{\sqrt{\Tr(J_i^\dagger J_i) \Tr(J_j^\dagger J_j)}},
\end{equation}
which directly defines the weight of the edge connecting the two vertices, capturing the degree of non-orthogonality.

Here, we provide a series of graph models corresponding to distinct optimally symmetric POVM as shown in the main text Fig. 2. Each site represents different channels and the edges that connecting them represent the orthogonality between each polarization channels. The thickness of the edge clearly shows the relation between the two polarization channels and would vanish when they are orthogonal to each other. The figures below each graph illustrate the corresponding polarization state in the Bloch sphere. Therefore, for given graph model, we could design the metasurface that functionate the polarization splitting state in analogy to the corresponding state in Bloch sphere. The state we finally prepared could link to the desired graph model by direct mapping, and can be used for the multiport random number generation.

As discussed above, we have designed the three component metasurface to generate six diffraction orders. For each design containing two diffraction orders, it serve as polarization analysers oriented along two orthogonal directions. The corresponding analyser matrices are $J_i=a_i\ \left.|\gamma_i\right\rangle\left\langle\gamma_i|\right.$, where $\gamma_{+1}=R\left(\phi\right)\left(\begin{matrix}\cos{(\chi)}\\-i\sin{(\chi)}\\\end{matrix}\right)$ is the polarization state and $a_i$ is the normalization coefficient.

\subsection{Security against coherent attacks}

In the analysis in main text, we derived the conditional min-entropy \( H_{\min}(A|E) \) for a single metasurface measurement round, assuming potential entanglement among photons within that round and Eve's external quantum memory. This bound, rooted in the POVM graph's weighted degrees, secures against non-collective attacks. To achieve composable security under general coherent attacks, where Eve entangles photons across multiple rounds, we examine \( k \) protocol repetitions. Eve may craft a \( k \)-mode entangled state \( \rho^{(mk)} \) to optimize her joint guessing probability over all \( k \) outcomes.

Now consider \(k\) sequential rounds of the metasurface. In each round \(r=1,\dots,k\), Alice again measures \(m\) photons with the same POVM \(\{F_i\}_{i=1}^N\). Let \(\mathbf{i}=(i_1,\dots,i_k)\) denote the string of outcomes, and define the corresponding POVM element on the \(km\)-photon system by
\begin{equation}
	F_{\mathbf{i}}=	F_{i_1}^{\otimes m} \otimes \cdots \otimes F_{i_k}^{\otimes m}.\nonumber
\end{equation}
Eve performs an arbitrary POVM \(\{M_{\mathbf{i}}^E\}\) on her quantum memory to guess the entire string \(\mathbf{i}\). Allowing the most general coherent attack over all \(km\) photons and Eve's memory, the optimal guessing probability is
\begin{eqnarray}
	p_{\mathrm{guess}}^{(k)}(A^k|\rho_{A}^{mk}, F_{\mathbf{i}})=\max_{\{\{M^{E}_{\mathbf{i}}\},\,\ket{\psi_{A^{k}E}}\}}	\sum_{\mathbf{i}}\Tr\!\left[F_{\mathbf{i}} \otimes M^{E}_{\mathbf{i}}\,	\ketbra{\psi_{A^{k}E}}\right].	\nonumber
\end{eqnarray}
From \(F_{\mathbf{i}}\otimes I_E \le \lambda_{\max}(F_{\mathbf{i}})\,I_{A^{k}E} \) ($I$ is the fully mixed state), as the guessing probability is the maximum, over all input states that Eve can prepare, of her success probability for the given POVM, we obtain
\begin{align}
	p_{\mathrm{guess}}^{(k)}(A^k|\rho_{A}^{mk}, F_{\mathbf{i}})
	&=
	\begin{aligned}[t]
		\max_{\mathbf{i}}
		\max_{\{\{M^{E}_{\mathbf{i}}\},\,\ket{\psi_{A^{k}E}}\}}
		\Tr\!\Big[
		\bigl(\lambda_{\max}(F_{\mathbf{i}})I_{A^{k}E}\bigr)
		(I_{A^{k}} \otimes M^{E}_{\mathbf{i}})
		\ketbra{\psi_{A^{k}E}}
		\Big]
	\end{aligned}\nonumber\\
	&=
	\begin{aligned}[t]
		\max_{i_1,\dots,i_k}	\prod_{r=1}^{k}
		\max_{\{\{M^{E}_{i_r}\},\,\ket{\psi_{AE}}\}}
		\Tr\!\Big[
		\bigl(\lambda_{\max}\!\bigl(F_{i_r}^{\otimes m}\bigr)I_{AE}\bigr)
		(I_{A} \otimes M^{E}_{i_r})
		\ketbra{\psi_{AE}}
		\Big]
	\end{aligned}\nonumber\\
	&=
	\bigl(p_{\mathrm{guess}}(A|\rho_{A}^{m}, \{F_{i}\}_{N})\bigr)^k,
	\label{eq:pguess_km_upper_unified}
\end{align}
where $ \lambda_{\max}(\cdot) $ is the largest eigenvalue. The second equality follows from the fact that the maximum is attained when the state is prepared as the projector onto the eigenvector corresponding to the largest eigenvalue of the POVM element. Because the POVM \(\{F_{\mathbf{i}}\}\) has a tensor‑product structure, the corresponding eigenvector factorizes, and thus the optimal state is a tensor‑product state. Thus, the optimal coherent attack is achieved by a fully separable attack across rounds: in every round \(r\), Eve prepares on Alice's \(m\) photons the same single-round optimal state \(\sigma_{A i^\star}=\Tr_E \!\Big[(I_{A} \otimes M^{E}_{i^\star})\ketbra{\psi^\star_{AE}}\Big]/p_{i^\star}\) ($p_{i^\star}$ is the normalized parameter) and always guesses the string \(\mathbf{i}^\star=(i^\star,\dots,i^\star)\). Hence the optimal attack is effectively a product across different rounds, and we have
\begin{equation}
	H_{\min}^{(k)}(A^k|E)=k\,H_{\min}(A|E).
	\nonumber
\end{equation}

In fact, our single-use analysis for \(m\) photons already allowed \(\rho_A^m\) to be an arbitrary state, potentially maximally entangled across all measured photons and with Eve's memory. If we now treat the \(k\) sequential rounds as a single ``super round'' in which Alice measures \(km\) photons with the tensor-product POVM \(\{F_{\mathbf{i}}\}\), Eve's most general coherent attack is simply an arbitrary state on these \(km\) photons and her quantum memory, exactly the same setting as in the single-use analysis, with \(m\) replaced by \(km\). Since the bound on \(p_{\mathrm{guess}}\) depends only on the POVM operators and not on the structure of the input state, it applies verbatim to this \(km\)-photon POVM. Consequently, our graph-theoretic bound on the single-round guessing probability extends straightforwardly to \(k\) uses and guarantees composable security against arbitrary coherent attacks.

\subsection{Randomness estimation with the extractor}

In the randomness extraction process using a $\textit{universal}_2$ hash function, the final output sequence may deviate from an ideal uniform distribution. This deviation is rigorously bounded by the quantum leftover hash lemma [44]. Let $\rho'_{A^{k}E}$ denote the joint state of the extracted sequence $A^{k}$ and Eve’s system $E$ after hashing. The $\Delta$-security of the randomness sequence—meaning its statistical distance from a perfectly uniform and independent state—is defined as
\begin{eqnarray}
	\min_{\sigma_{E}}\frac{1}{2}\parallel \rho'_{A^{k}E}-\omega_{A^{k}}\otimes\sigma_{E} \parallel_{1}\leq \Delta\nonumber
\end{eqnarray}
where $\omega_{A^{k}}$ is the fully mixed state, and $\sigma_E$ is an arbitrary state held by Eve. By the quantum leftover hash lemma, a randomly chosen $\textit{universal}_2$ hash function enables extraction of a $\Delta$-secure sequence of length $l$, satisfying
\begin{eqnarray}
	\Delta\leq\frac{1}{2}\times2^{\sqrt{l-H^{(k)}_{\min}({A^{k}}|E)}}.\nonumber
\end{eqnarray}
To account for extraction failure, we set the failure probability to $\varepsilon \geq \Delta$. The final extractable certified randomness per $m$-photon event, constrained by the weighted degrees of the POVM graph, is thus bounded by:
\begin{eqnarray}
	l\geq k\left (m\log_{2}\left (\min_{i}D_{i}\right )-\log_{2}g_{m}\right )-2\log_{2}\frac{1}{2\varepsilon}.\nonumber
\end{eqnarray}
Similarly, for a mixed Fock-state source, the total extractable randomness per pulse is
\begin{eqnarray}
	l\geq k\left (\sum_{m=1}^{N-1}\gamma_{m}m\log_{2}\left (\min_{i}D_{i}\right )-\log_{2}g_{m}\right )-2\log_{2}\frac{1}{2\varepsilon}.\nonumber
\end{eqnarray}
In our experiment, we adopt a conservative security parameter of $\varepsilon = 2^{-50}$, ensuring privacy for the extracted bitstream across both heralded single-photon and pulsed laser sources.

\subsection{Traditional QRNG estimation}

In conventional QRNG estimation, the source is assumed to be a trusted and well-characterized pure state, and no adversarial attack on the device is considered. For a known input state $\ket{\psi}$ and a POVM $\{F_{i}\}_{N}$ describing the measurement, the output probabilities are
\begin{equation}
	p_{i} = \Tr\!\left(F_{i}\ketbra{\psi}{\psi}\right).
\end{equation}
The conditional min-entropy of the outcome (conditioned only on the classical description of the device and the state) is then estimated via the maximum output probability,
\begin{equation}
	H_{\min}^{\mathrm{(trad)}}(A) = -\log_{2}\bigl(\max_{i} p_{i}\bigr),
\end{equation}
which corresponds to the adversary’s optimal guessing probability in this trusted scenario,
\begin{equation}
	p_{\text{guess}}^{\mathrm{(trad)}}(A) = \max_{i} p_{i}.
\end{equation}
This ``traditional'' estimation is valid only under the assumption that the adversary does not interact with the device and that the input state is not entangled with any external system.

However, if an adversary can prepare or influence the input state, this approach is no longer secure, as discussed in the main text. For example, suppose the actual input is one photon from an entangled pair, so that the reduced state on the device side is a completely mixed state. In this case, although the local statistics may appear uniform, the adversary can possess quantum side information (the other half of the entangled pair) and thus can correlate her measurement with the device outcomes. The simple bound based on $\max_{i} p_{i}$ then fails to capture the true guessing probability and cannot guarantee security. In such an adversarial setting, one must instead bound the conditional guessing probability with respect to arbitrary quantum side information. As discussed above, the corresponding min-entropy must be bounded using some new methods, such as our graph-theoretic method, rather than the traditional $\max$-probability estimate.

\subsection{Mixed input states with entanglement dependence security}

In this note we give a more detailed and explicit analysis of why mixed input states can lead to a complete loss of security when the source is correlated or entangled with an adversary, even though the local output statistics remain unchanged. The probability-based QRNG estimation with traditional QRNG estimation assigns a randomness rate to each observed distribution \(\{p_i\}\) (e.g.\ the blue solid curves in Fig.~S4(a--f)), ignoring any side information and treating the device as if it were isolated. In contrast, the true device-independent security is characterized by the guessing probability, which can be strictly larger than the value suggested by that probability-only estimate.

As shown in the main text Fig.~S4(c, f), we first consider the case where the input state is completely mixed: $	\rho_A = \frac{I}{2}, ~ (|s| = 0)$. A natural way this can arise is when the device receives one photon of a maximally entangled pair, e.g.\ the Bell state
\begin{equation}
	\ket{\Phi^+}_{AE}
	= \frac{1}{\sqrt{2}}\bigl(\ket{H}_A\ket{H}_E + \ket{V}_A\ket{V}_E\bigr).
	\label{eq:Bell}
\end{equation}
Tracing out Eve’s system yields
\begin{equation}
	\rho_A = \operatorname{Tr}_E\!\bigl[\ket{\Phi^+}\!\bra{\Phi^+}_{AE}\bigr]
	= \frac{1}{2}\bigl(\ket{H}\!\bra{H} + \ket{V}\!\bra{V}\bigr)
	= \frac{I}{2}.
	\label{eq:rhoA-mixed}
\end{equation}

The completely mixed state admits infinitely many ensemble decompositions. Two
relevant examples are:
\begin{align}
	\rho_A
	&= \frac{1}{2}\bigl(\ket{H}\!\bra{H} + \ket{V}\!\bra{V}\bigr),
	\label{eq:HV-decomp}\\[1ex]
	&= \frac{1}{2}\bigl(\ket{L}\!\bra{L} + \ket{R}\!\bra{R}\bigr),
	\label{eq:LR-decomp}
\end{align}
where \(\ket{L},\ket{R}\) denote left- and right-circular polarization states. These are precisely the types of decompositions we refer to as the HV and LR decompositions in Fig.~S4(c, f). By the Hughston–Jozsa–Wootters (HJW) theorem, any ensemble decomposition of \(\rho_A\) can be realized by performing a suitable measurement on system \(E\) in a fixed purification \(\ket{\Psi}_{AE}\). Concretely, for the Bell state Eq. \eqref{eq:Bell}:
\begin{itemize}
	\item If Eve measures in the \(\{\ket{H},\ket{V}\}\) basis on \(E\), then conditional on her outcome she steers system \(A\) to either \(\ket{H}\) or \(\ket{V}\). This realizes the HV decomposition Eq. \eqref{eq:HV-decomp}.
	\item If Eve instead measures in the \(\{\ket{L},\ket{R}\}\) basis on \(E\), she steers \(A\) to \(\ket{L}\) or \(\ket{R}\), realizing the LR decomposition Eq. \eqref{eq:LR-decomp}.
\end{itemize}
Thus, although the reduced state at the device is always \(\rho_A= I/2\), Eve can choose—by changing her local measurement basis \(M^E\)—which decomposition (HV, LR, or any other) is effectively realized in each round.

Crucially, the device’s observed probabilities \(p_{i} = \operatorname{Tr}[F_{i} \rho_A]\) depend only on \(\rho_A\) and the POVM \(\{F_{i}\}_{N}\), and therefore are independent of Eve’s choice of decomposition or measurement on \(E\). For \(|s|=0\), the local distribution \(\{p_{i}\}\) is completely fixed and gives a single probability-based randomness value (blue solid curve in Fig.~S4(c, f)).

However, Eve’s optimal guessing probability is very different for different decompositions. This is exactly what is plotted in Fig.~S4(c, f): the green and orange dashed curves show the larger guessing probabilities that Eve can achieve when she is correlated with a specific decomposition (HV or LR), while the blue solid curve comes from the traditional estimation that only uses the local probabilities. The gap between these dashed and solid curves is the “unsecure gap”: the probability-based method overestimates the amount of secure randomness because it completely ignores quantum side information and decomposition dependence.
\\
\hfill

\begin{figure*}[htbp]
	\centering
	\includegraphics[width=6.5in]{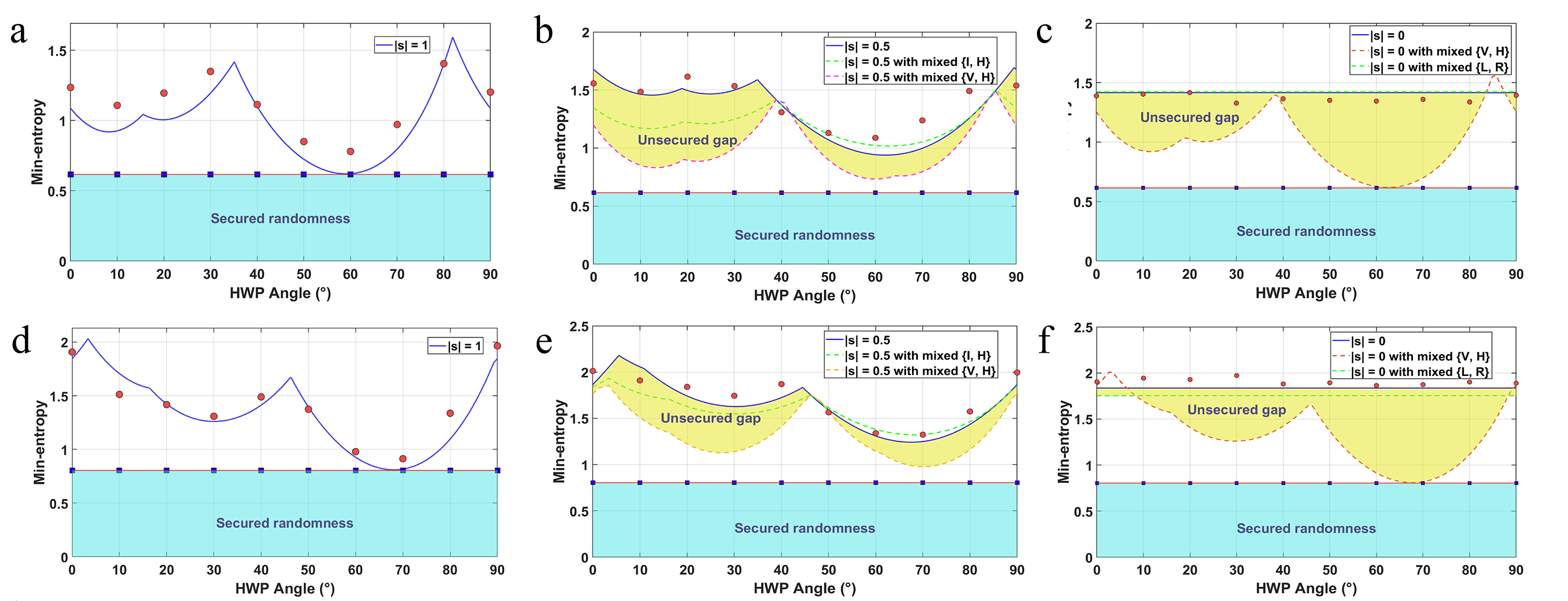}
	\caption{Min-entropy as a function of the input state for the 4‑channel (a--c) and 6‑channel (d--f) metasurfaces, comparing the graph-theoretic bound (light green shaded region) with the randomness inferred directly from output probabilities (blue line: simulation; red points: experiment). (a, d): pure H-polarized input with varying HWP angle, showing strong state dependence of probability-based randomness in contrast to the polarization-independent graph-theoretic bound. (b, c, e, f): mixed polarization inputs with Bloch-vector modulus $\abs{s}=0.5 $ (b, e) and $\abs{s}=0 $ (c, f), scanned via HWP angle; probability-based randomness overestimates the secure randomness because an adversary correlated with the chosen decomposition (green and orange dashed lines) can gain extra information, whereas the graph-theoretic bound remains secure and state independent.}
	\label{fig:ent}
\end{figure*}

We now turn to the case where the Bloch vector of the input has modulus \(|s|=0.5\). Assuming \(s\) points along the \(H/V\) axis, we can write $	\rho_A = \frac{1}{2}(I + s \sigma_z)$ $(|s| = 0.5$), with eigenvalues
\begin{equation}
	\lambda_H = \frac{1+s}{2} = 0.75, \qquad
	\lambda_V = \frac{1-s}{2} = 0.25.
\end{equation}
Thus
\begin{equation}
	\rho_A = 0.75\,\ket{H}\!\bra{H} + 0.25\,\ket{V}\!\bra{V}.
	\label{eq:rhoA-HV}
\end{equation}
Equivalently, \(\rho_A\) can be seen as a mixture of a maximally mixed state and a pure state:
\begin{equation}
	\rho_A
	= (1-\lambda)\,\frac{I}{2} + \lambda\,\ket{H}\!\bra{H},
	\qquad \lambda = 0.5.
	\label{eq:rhoA-IH}
\end{equation}
In this sense, \(\rho_A\) can be viewed as “a partially mixed state (\(I/2\)) plus a pure \(\ket{H}\) component”. A simple purification of \(\rho_A\) is given by a partially entangled state
\begin{equation}
	\ket{\Psi}_{AE}
	= \sqrt{0.75}\,\ket{H}_A\ket{0}_E
	+ \sqrt{0.25}\,\ket{V}_A\ket{1}_E,
	\label{eq:partially-entangled}
\end{equation}
for which
\begin{equation}
	\operatorname{Tr}_E\!\bigl[\ket{\Psi}\!\bra{\Psi}_{AE}\bigr] = \rho_A.
\end{equation}
Again, Eve can choose different measurements on \(E\) to realize different ensemble decompositions of \(\rho_A\). The spectral decomposition Eq. \eqref{eq:rhoA-HV} corresponds to an “HV decomposition”. However, there exist many other decompositions, for instance into states that involve \(\ket{V}\), \(\ket{H}\), and some intermediate polarizations like \(I\) or \(\ket{D},\ket{A}\). In particular, one can find ensembles of the form
\begin{equation}
	\rho_A = \sum_k q_k  \tau_k ,
	\label{eq:general-decomp}
\end{equation}
where \(\{\tau_k\}\) are different polarization states (e.g.\ VH or $I$H decompositions as used in Fig.~S4(b, e)), and \(\{q_k\}\) is a probability distribution. By the HJW theorem, each such ensemble Eq. \eqref{eq:general-decomp} corresponds to some POVM on Eve’s system \(E\), i.e.\ to some choice of measurement basis \(M^E\). Thus, for \(|s|=0.5\) the same density matrix \(\rho_A\) and hence the same local statistics \(\{p_{i}\}\) (for each HWP angle) are compatible with many different adversarial strategies, depending on how Eve chooses to measure her system and thereby which decomposition she effectively realizes.

In Fig.~S4(b, e), the green and orange dashed curves represent the randomness for two such adversarial strategies, corresponding to different decompositions (VH, $I$H) of the same \(\rho_A\). Both dashed curves lie above the probability-based estimate (blue solid curve), demonstrating that the probability-only method again overestimates the secure randomness. Moreover, the fact that the green and orange dashed curves differ from each other highlights the decomposition dependence: for the same local statistics and the same \(|s|=0.5\), different choices of decomposition (i.e.\ different measurements on Eve’s system) lead to different amounts of information for Eve.
\\
\hfill

The fundamental issue revealed above is that local statistics \(\{p_{i}\}\) and even the density matrix \(\rho_A\) do not uniquely specify the underlying ensemble or purification when an adversarial source is allowed. Different decompositions of the same \(\rho_A\) are indistinguishable at the level of observed probabilities but can correspond to very different guessing probability. Any security analysis that relies on a particular source model or on a fixed decomposition is therefore not robust in the presence of a malicious or untrusted source.

Our graph-theoretic bound addresses this by being explicitly source independent. Operationally, the interplay between \(\{p_{i}\}\) and the structure of graph allows us to derive a graph-theoretic quantity that bounds Eve’s optimal guessing probability in the worst case, and is valid for \emph{all} decompositions and purifications compatible with the experiment. This leads to a corresponding lower bound on the secure randomness rate that is independent of any particular source model or decomposition and closes the unsecure gap exhibited by the probability-based estimation. In particular, for both \(|s|=0\) (completely mixed inputs) and \(|s|=0.5\) (partially mixed inputs), our graph-theoretic bound correctly accounts for the worst-case quantum side information that Eve may possess (e.g.\ due to entanglement with the source), and thus guarantees security even when the source is fully under Eve’s control.

\subsection{Metasurface characterizations}

We characterized the fabricated TiO$_{2}$ metasurfaces using a continuous-wave laser at 808 nm, as shown in Fig. \ref{fig:expcoh}. A combination of a half-wave plate (HWP) and a quarter-wave plate (QWP) placed before the sample generated six well-defined input polarization states spanning the Poincaré sphere. For each input state and each diffraction order, we recorded the output power enabling complete reconstruction of the Jones matrix of every channel.

The experimentally retrieved, normalized transmission matrices for the $N = 6$ and $N = 4$ devices are
\begin{eqnarray}
	&&T_{6} = \begin{bmatrix}
		1.0000 & 0.5611 + 0.2397i \\
		0.1992 + 0.5189i & 0.7312 + 0.0255i \\
		0.4675 - 0.3481i & 1.2228 - 0.5759i \\
		-1.0536 + 0.1945i & 0.3723 - 0.0072i \\
		0.5198 + 0.0852i & 0.1845 + 0.5752i \\
		0.2995 - 0.8241i & -0.7946 + 1.0647i
	\end{bmatrix}
	\nonumber\\&&T_{4} = \begin{bmatrix}
		1.0000           & -1.7137 + 0.2360i \\
		0.4383 + 0.8435i & -0.4219 - 0.7904i \\
		-0.7775 - 0.3838i & -1.2293 - 0.4971i \\
		-0.9788 + 0.9508i & -0.2429 + 0.7208i
	\end{bmatrix}
\end{eqnarray}

To provide a more intuitive geometric interpretation of the measurement bases realized by each output port, we represent the experimentally reconstructed Jones matrices $J_i$ in terms of their equivalent Bloch vectors. Each port implements a non-ideal polarization projection that can be expressed as the POVM element
$$J_i = \frac{1}{2} \bigl( I + \boldsymbol{v}_i \cdot \boldsymbol{\sigma} \bigr),$$
where $\boldsymbol{\sigma} = (\sigma_x, \sigma_y, \sigma_z)$ is the vector of Pauli matrices, and $\boldsymbol{v}_i = (v_{i,x}, v_{i,y}, v_{i,z})$ is the Bloch vector of the $i$-th channel (given by the $i$-th row of the matrix $V$). The magnitude $|\boldsymbol{v}_i| < 1$ quantifies the polarization purity influenced by fabrication imperfections in that channel, while the direction of $\boldsymbol{v}_i$ indicates the eigenpolarization toward which the projection is biased.

The experimentally extracted Bloch-vector matrix for the six (N = 6) and four (N = 4) output ports are
\begin{eqnarray}
	V_{6} = \begin{bmatrix}
		0.9506 & 0.3032 & 0.0667 \\
		0.3760 & -0.8869 & -0.2682 \\
		0.8045 & 0.1535 & -0.5738 \\
		-0.7304 & -0.0898 & 0.6771 \\
		0.4403 & 0.8939 & -0.0846 \\
		-0.9704 & -0.2126 & -0.1146
	\end{bmatrix}\qquad
	V_{4} = \begin{bmatrix}
		-0.8585 &  0.1182 & -0.4990 \\
		-0.9982 &  0.0111 &  0.0591 \\
		0.9136 & -0.0680 & -0.4010 \\
		0.7564 & -0.3888 &  0.5259
	\end{bmatrix}
\end{eqnarray}
with each row corresponding to one output port (visualized on the Poincaré sphere in Fig. 2 of the main text). The measured relative diffraction efficiencies are
$$\boldsymbol{\eta}_6 = (0.177,\; 0.097,\; 0.224,\; 0.153,\; 0.068,\; 0.280),
\qquad
\boldsymbol{\eta}_4 = (0.375,\; 0.160,\; 0.236,\; 0.229).$$ 
Far-field images and wavelength-dependent efficiency curves confirm that $>$85 \% of the incident power is directed into the designed diffraction orders across a $\sim$100 nm bandwidth, with residual zero-order light arising from minor fabrication imperfections that are fully accounted for in the calibration.

These experimentally calibrated Jones matrices and efficiencies serve as the exact POVM elements $F_i = t p_i J_i$ used in all subsequent randomness analyses, ensuring that the theoretical security bounds derived from the graph-theoretic framework are directly applicable to the physical device.

\subsection{Heralded single photon measurements with the metasurface}

In our experiment, herald single photon pairs are generated via type-I spontaneous parametric down-conversion (SPDC) in a 0.5 mm-thick $\beta$-barium borate (BBO) crystal. The crystal is pumped by a vertically polarized, frequency-doubled 404 nm pulsed laser ($\sim$140 fs pulse duration, 80 MHz repetition rate). The down-converted signal and idler photons are produced in a noncollinear, degenerate configuration at 808 nm with orthogonal horizontal (H) linear polarizations. After the BBO crystal, residual pump light is removed by a series of long-pass filters, and the photon pairs are spatially separated using a prism. The idler photon is directed to a fiber-coupled silicon avalanche photodiode serving as the heralding detector, while the signal photon is focused onto the metasurface under test. Photons diffracted into the designed output orders are collected by a 1-to-N fiber coupler array precisely aligned to the far-field pattern of the metasurface and routed to fiber-coupled single-photon detectors (SPDs). This heralded single-photon configuration ensures well-defined temporal arrival at the metasurface by recording the coincidence events.

\subsection{Time-division multiplexing of metasurface output ports}

\begin{figure*}[htbp]
	\centering
	\includegraphics[width=2in]{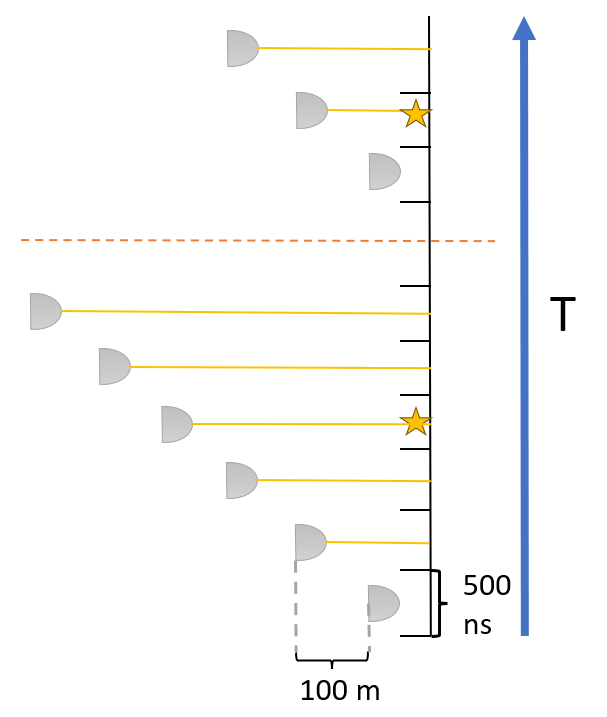}
	\caption{Schematic of the time-division multiplexing with a single SPD: 500-ns time windows are assigned to 4 or 6 diffraction channels using 100 m path-length differences in optical fibers.}
	\label{fig:tdm}
\end{figure*}

In many practical quantum information applications—particularly when using cryogenic detectors such as SPDs, the cost and complexity associated with large detector arrays can become prohibitive. To circumvent this limitation while preserving the full N-port POVM structure of the metasurface, we implement temporal multiplexing of the diffracted channels, as shown in Fig. \ref{fig:tdm}.

Each output port is coupled into a single-mode fiber and routed through a different optical delay line realized with fiber lengths of approximately 100 m, 200 m, 300 m, 400 m, and 500 m (for the N = 6 device; correspondingly shorter lengths are used for N = 4). These lengths produce well-separated temporal delays of roughly 500 ns, 1000 ns, 1500 ns, 2000 ns, and 2500 ns, respectively, yielding a total temporal window of ~2.5 µs. After the delay stage, all fibers are combined via a cascaded series of low-loss $1\times2$ fiber beam splitters into a single high-efficiency SPD with the detection probability $\delta=0.6$ and the dead time $\sim$440 ns.

A high-precision time-to-digital converter (TDC) synchronized to the pump laser records the arrival time of each detected photon with sub-nanosecond resolution. A simple time-windowing algorithm then unambiguously assigns every detection event to its originating metasurface port based on its absolute arrival time within the 2.5 µs frame. The resulting time-tagged event list is functionally equivalent to data acquired with N independent detectors, yet requires only one physical detector and one timing channel.

This temporal multiplexing scheme offers several key practical advantages: it dramatically reduces the required number of detectors from N to just one (along with the associated readout electronics), eliminates channel-to-channel variations in detection efficiency and timing jitter, scales straightforwardly to larger N simply by extending the fiber delay lengths, and relies entirely on commercially available fiber-optic components and standard telecom-grade timing hardware. The primary trade-off—a modest reduction in the maximum sustainable event rate imposed by the single detector’s dead time and the total frame duration—remains negligible at the heralded single-photon rates in our experiments ($<$ 10 kHz). Overall, this detector-count-agnostic approach enables efficient experimental implementation of high-dimensional, polarization-independent POVM measurements with metasurfaces without compromising scalability or performance.

\subsection{Mixed Fock-state (coherent-state) measurements with the metasurface}

\begin{figure*}[htbp]
	\centering
	\includegraphics[width=5in]{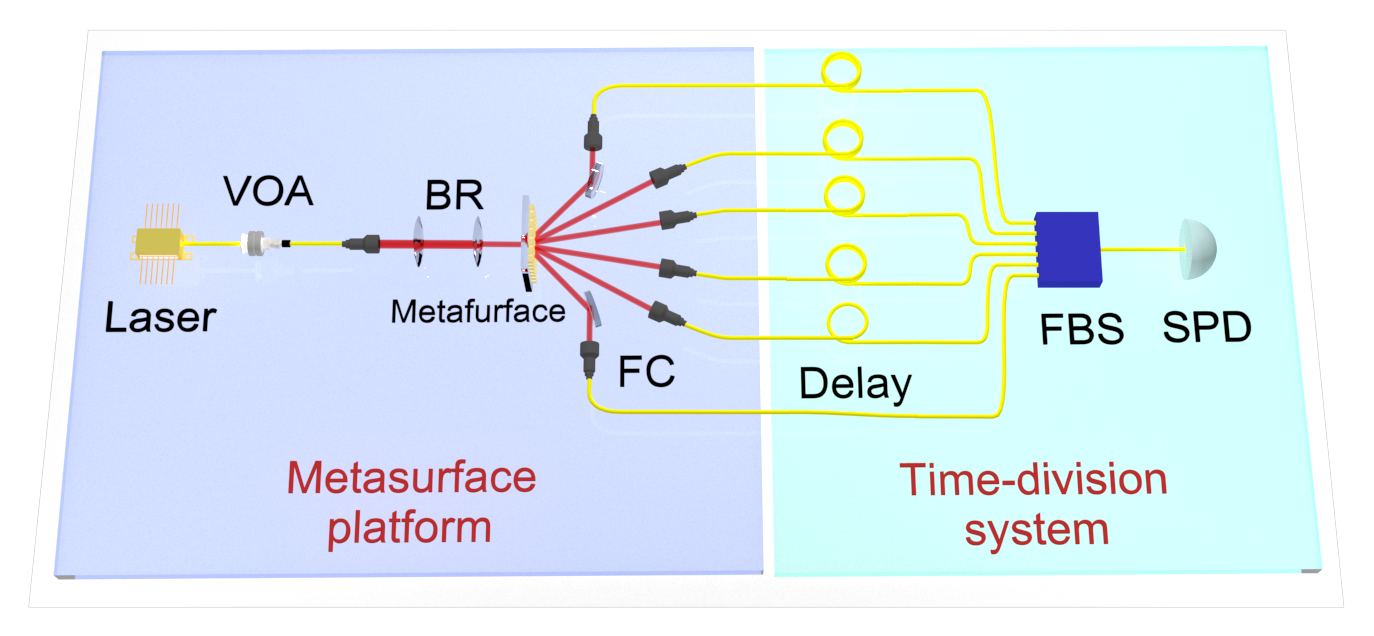}
	\caption{Experimental setup for coherent-state (mixed Fock-state) measurements and classical characterization of the TiO$_{2}$ metasurfaces. The laser beam is first conditioned by a beam reducer (BR) and then attenuated by a variable optical attenuator (VOA) to set the mean photon number at the metasurface. The diffracted light in the designed output channels is coupled into optical fibers by fiber couplers (FC) and analyzed using a fiber beam splitter (FBS) and fiber-coupled single-photon detectors (SPDs).}
	\label{fig:expcoh}
\end{figure*}

The experimental configuration for the mixed Fock-state measurements is shown in Fig.~\ref{fig:expcoh}. In this case, the SPDC source is replaced by a pulsed laser operating at a repetition rate of 300 kHz, with central wavelength matched to the design wavelength of the metasurface (808 nm in our experiment). The laser output is first cleaned spatially and spectrally, and then sent through a variable optical attenuator (VOA), which precisely controls the mean photon number per pulse at the metasurface plane. The attenuated beam is focused onto the metasurface under test with the same illumination geometry as in the heralded single-photon measurements. Photons diffracted into the designed output orders are collected by the 1-to-$N$ fiber-coupler array aligned to the far-field diffraction pattern and routed to fiber-coupled SPDs, identical to the single-photon configuration.

Each pulse from the attenuated laser can be modeled as a weak coherent state $\lvert \alpha \rangle$ with amplitude $\alpha = \sqrt{\mu}\,e^{i\phi}$ and mean photon number $\mu = \lvert \alpha \rvert^{2}$. In the Fock basis, a coherent state can be expanded as
\begin{equation}
	\lvert \alpha \rangle 
	= e^{-\mu/2} \sum_{n=0}^{\infty} \frac{\alpha^n}{\sqrt{n!}} \lvert n \rangle .
\end{equation}
For the general pulsed laser, the output phase $\phi$ of each pulse varies from 0 to 2$\pi$; therefore, the relevant output state is obtained by averaging over the global phase,
\begin{equation}
	\rho(\mu) 
	= \frac{1}{2\pi} \int_{0}^{2\pi} \mathrm{d}\phi \,
	\bigl\lvert \sqrt{\mu}\,e^{i\phi} \bigr\rangle 
	\bigl\langle \sqrt{\mu}\,e^{i\phi} \bigr\rvert .
\end{equation}
Substituting the Fock expansion into the above expression, the phase integral removes all off-diagonal terms and yields
\begin{equation}
	\rho(\mu) 
	= \sum_{n=0}^{\infty} P_n(\mu) \lvert n \rangle \langle n \rvert ,
	\qquad
	P_n(\mu) = e^{-\mu} \frac{\mu^n}{n!} ,
\end{equation}
i.e., an effective mixed Fock state described by a classical Poisson photon-number distribution. The VOA is adjusted such that the mean photon number per pulse is in the single-photon regime, with an optimal operating point around $\mu \approx 0.4$ photons per pulse (as discussed in the main text and shown in Fig.~4(e)), which balances the randomness generation rate against the probability of multi-photon events.

Compared to the heralded single-photon configuration, the mixed Fock source sacrifices the strict single-photon character and thus reduces the conditional min-entropy due to the presence of multi-photon components. However, it offers important practical advantages: a simplified optical setup without nonlinear crystals or heralding detectors, operation at a high and easily tunable repetition rate, and straightforward integration with compact laser modules. These features make the mixed Fock-state configuration particularly attractive for scalable, practical QRNG implementations based on metasurfaces.

\subsection{Randomness extraction results}

Detection events from each metasurface port were assigned unique binary labels: 2-bit labels (00, 01, 10, 11) for the N=4 device and 3-bit labels (000, ..., 111) for the N=6 device, thereby directly mapping spatial diffraction channels to raw bit strings. Secure randomness was extracted offline using a Toeplitz-matrix extractor, and the output block length was conservatively chosen according to the composable conditional min-entropy bounds derived from our graph-theoretic framework and experimentally calibrated channel efficiencies.

With the heralded single-photon source, the measured certified randomness rates are 0.6153 secure bits ($N=4$) and 0.8059 secure bits ($N=6$) per detected photon, which demonstrates scaling with the number of non-orthogonal ports. When switching to the more practical pulsed-laser direct illumination (Poisson-distributed photon statistics, mean photon number $\sim$0.4), the extracted randomness per laser pulse drops because of residual multi-photon contributions. Nevertheless, both devices still deliver finite certified randomness 0.1192 secure bits ($N=4$) and 0.1220 secure bits ($N=6$) per detected photon. Even with an untrusted classical laser source, the metasurface platform therefore maintains a positive, composably secure randomness rate suitable for real-world QRNG applications.

\subsection{Randomness test}

\begin{figure*}[htbp]
	\centering
	\subfigure[]{\includegraphics[width=3in]{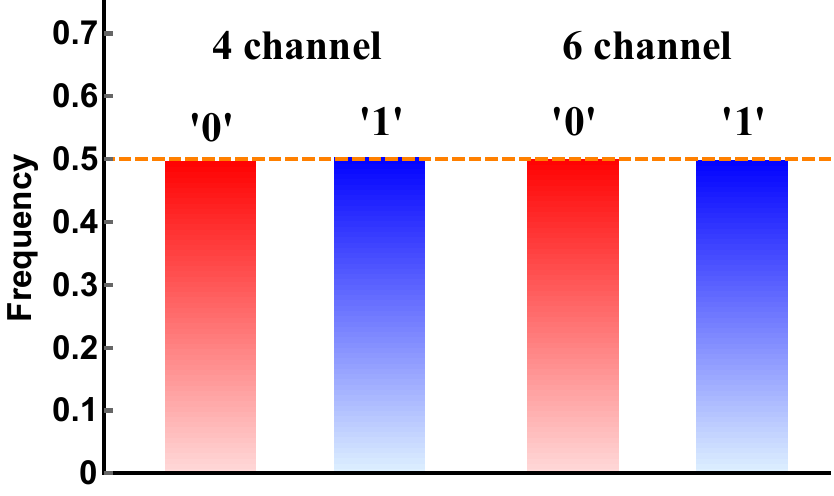}}\label{fre}
	\hspace{0.6in} 
	\subfigure[]{\includegraphics[width=3in]{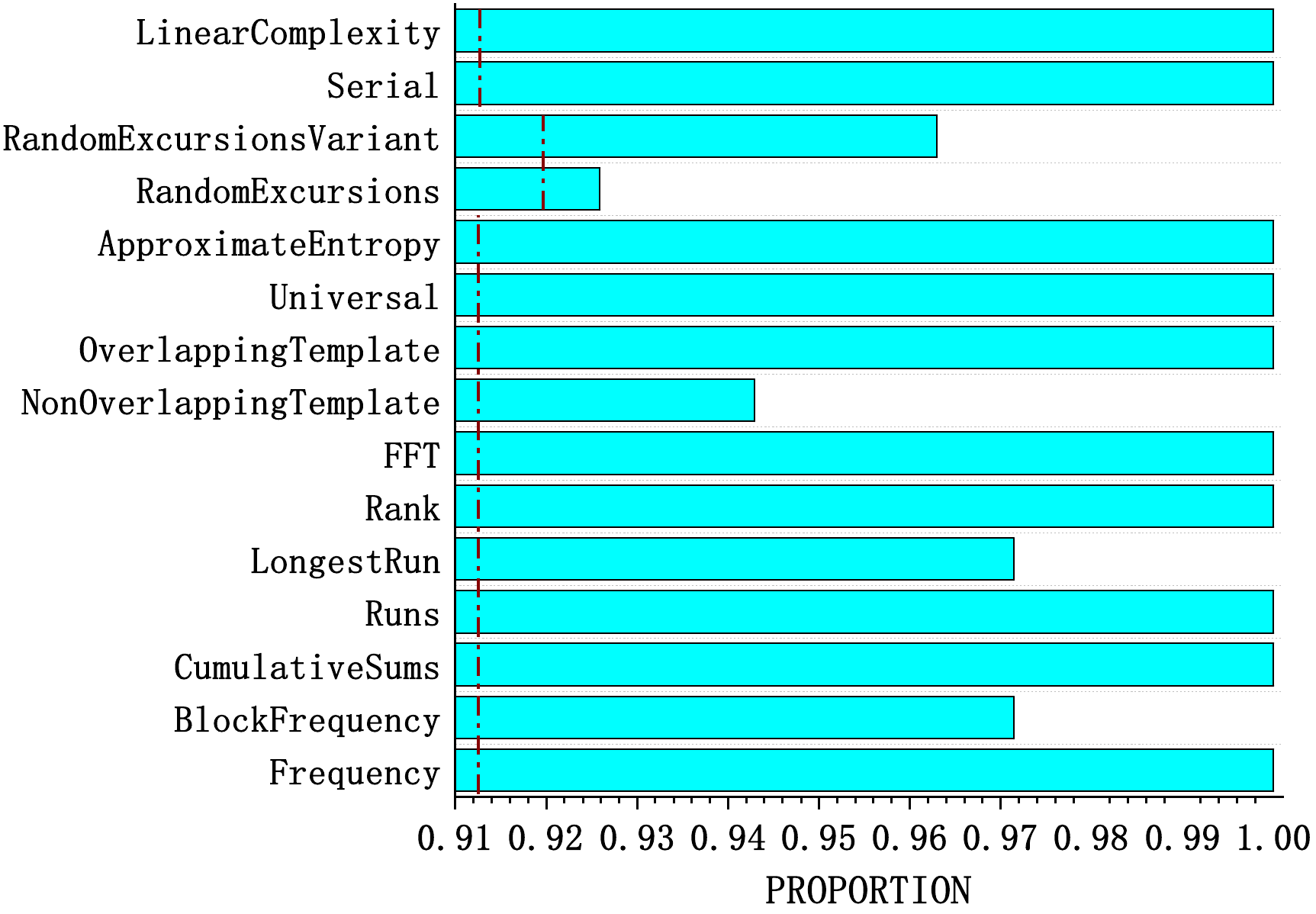}}\label{nistsfpro}
	\hspace{0.3in} 
	\subfigure[]{\includegraphics[width=3in]{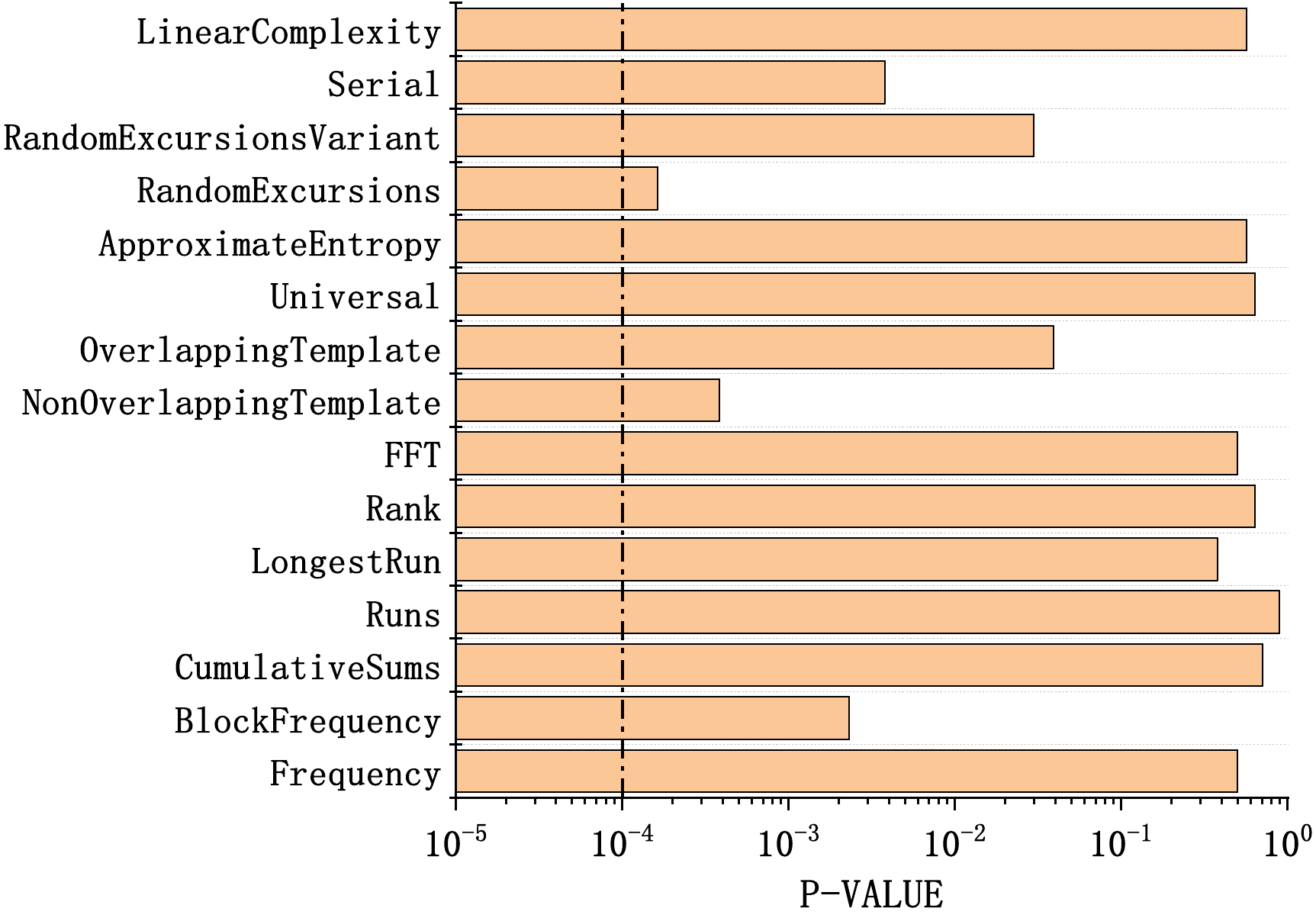}}\label{nistsfp}
	\caption{ (a) Final bit frequency distribution (probability of “0” and “1”) in the extracted random sequences for the 4-channel and 6-channel single photon systems, compared to the ideal uniform value of 0.5. (b) and (c) show the results of the NIST statistical test suite with different terms of P-value$_T$ and proportion for a coherent source input into the 6-channel metasurface.. The black dotted line is the passing line. All of the test items are passed. }
	\label{fig:nistsf}
\end{figure*}

For the 4-channel and 6-channel single-photon systems, the final bit-frequency distribution in the extracted random sequences is shown in Fig. \ref{fig:nistsf}(a). The results are close to the ideal uniform value of 0.5. Furthermore, to perform a more comprehensive statistical quality assessment using a longer sequence generated by a coherent source input into the 6-channel metasurface, we evaluated the final extracted bit streams with the full NIST SP 800-22 test suite [39]. For each metasurface configuration, we generated a single continuous bit stream comprising 35 independent sequences of 1 Mbit each. A test was passed if its P-value fell within the acceptance range $ 0.01\leq$ P-value $\leq0.99 $. The proportion of sequences passing each individual sub-test, together with the uniformity P-value of the P-values (P-value$_T$), were analyzed at a significance level $ \alpha=0.01 $.

As shown in Fig. \ref{fig:nistsf}(b), the pass proportions for all sub-tests (including multiple non-overlapping template tests with different patterns) lie well within the confidence interval (black dashed lines). The uniformity of the P-value distribution is confirmed in Fig. \ref{fig:nistsf}(c), where the lowest observed P-value$_T$ exceeds the NIST threshold of 0.0001. Even the most stringent sub-tests (e.g., longest runs of ones, binary matrix rank, and the full set of template tests) exhibit excellent performance.

While passing NIST tests does not constitute a formal proof of quantum randomness, the complete success across all metrics provides strong empirical evidence that output is statistically indistinguishable from an ideal uniform random source. These results, combined with the composable security analysis, confirm the practical reliability and cryptographic-grade quality of the randomness generated by our metasurface-based QRNG.

%\begin{figure*}[htbp]
%	\centering
%	\includegraphics[width=5in]{g}
%	\caption{Schematic illustration of valid and invalid detection events for the $m=2$ case. Because photons are indistinguishable bosons and the detectors are insensitive to photon number, only runs in which two distinct output ports fire are accepted as valid two-photon responses, and this post-selection generalizes to requiring exactly $m$ distinct ports for an $m$-photon input.}
%	\label{fig:g}
%\end{figure*}

%\begin{figure*}[htbp]
%	\centering
%	\includegraphics[width=9in]{data-1}
%	\caption{}
%	\label{fig:channel}
%\end{figure*}

\end{document}